\documentstyle[12pt,epsf,psfig,rotate]{article}
\input FEYNMAN 
\newcommand{\pr}{\paragraph{}}
\newcommand{\nn}{\nonumber}
\newcommand{\be}{\begin{equation}}
\newcommand{\ee}{\end{equation}}
\newcommand{\bea}{\begin{eqnarray}}
\newcommand{\eea}{\end{eqnarray}}
\newcommand{\benn}{\begin{eqnarray*}}
\newcommand{\eenn}{\end{eqnarray*}}

\begin{document}
\newcommand{\nd}[1]{/\hspace{-0.5em} #1}
\begin{titlepage}

\begin{flushright}
NTUA--72/98 \\
OUTP-98--44P \\
hep-lat/9806029 \\
\end{flushright}

\begin{centering}
\vspace{.05in}
{\Large {\bf Phase Structure of lattice $SU(2) \otimes U_S(1)$ 
three-dimensional Gauge Theory \\}}
 
\vspace{.4in}
{\bf  K. Farakos$^{a}$}, {\bf N.E. Mavromatos$^{b,\diamond}$} and 
{\bf D.  McNeill$^{b}$} \\

\vspace{.3in}
{\bf Abstract} \\
\vspace{.05in}
\end{centering}
{\small We discuss a phase diagram for 
a relativistic 
 $SU(2) \times U_{S}(1)$ 
lattice 
gauge theory, with emphasis on  
the
formation of a parity-invariant chiral condensate, 
in the case when the $U_{S}(1)$ field is
infinitely coupled, and the $SU(2)$ field is moved away from infinite
coupling by means of a strong-coupling expansion. 
We provide analytical arguments on 
the existence of (and partially derive) a critical line 
in coupling space, separating the phase of 
broken $SU(2)$ symmetry from that where the symmetry is unbroken. 
We review uncoventional (Kosterlitz-Thouless type)
superconducting properties of the model,
upon coupling it to external electromagnetic potentials.
We discuss 
the r\^ole of instantons of the unbroken subgroup $U(1) \in SU(2)$, 
in eventually
destroying superconductivity under certain circumstances. 
The model 
may have applications to the theory of high-temperature
superconductivity. In particular, we argue that 
in the regime of the couplings
leading to the broken $SU(2)$ phase, the model may provide an 
explanation on the appearance of a pseudo-gap phase, lying 
between the antiferromagnetic and the superconducting phases.
In such a phase,  
a fermion mass gap appears in the theory, but there is no 
phase coherence, due to 
the Kosterlitz-Thouless mode of symmetry breaking.
The absence of superconductivity in this phase is attributed
to  
non-perturbative effects (instantons) of 
the gauge field $U(1) \in SU(2)$.}

\vspace{0.5in}

\begin{flushleft} 

\vspace{.05in}
$^\diamond$P.P.A.R.C. Advanced Fellow. \\
$^{a}$National Technical University of Athens, Department of Physics,
Zografou Campus 157 73, Athens, Greece, \\
$^{b}$University of Oxford, Department of (Theoretical) Physics, 
1 Keble Road OX1 3NP, Oxford, U.K. \\
\vspace{.05in}
June 1998 \\
\end{flushleft} 

\end{titlepage}

\section{Introduction}
There has been a great deal of recent interest in the dynamical symmetry
breaking patterns of three-dimensional quantum gauge field theories, 
both from
the pure particle theory standpoint \cite{app,farak,kondo}, and as a tool
for describing 
models of High $T_c$ Superconductors \cite{dor,IanA,fm}. The
gauge theories studied in those works 
have been either $QED_3$ and variants of it~\cite{app,dor,IanA},
or $SU(2) \otimes 
U(1)$~\cite{farak,fm}.

{}From the condensed-matter view point, which motivated our 
approach to the subject, 
the key suggestion which lead to 
non-abelian dynamical gauge symmetry structure for the doped antiferromagnet,
was the {\it slave-fermion} spin-charge separation ansatz 
for physical electron operators 
at {\it each lattice site} $i$~\cite{fm}:
\be
\chi _{\alpha\beta,i} 
\equiv \left(
\begin{array}{cc}
c_1 \qquad c_2 \\
c_2^\dagger \qquad -c_1^\dagger \end{array}
\right)_i \equiv {\widehat \psi} _{\alpha\gamma,i}{\widehat z}_{\gamma\beta,i} =
\left(\begin{array}{cc}
\psi_1 \qquad \psi_2 \\
-\psi_2^\dagger \qquad \psi_1^\dagger \end{array}
\right)_i~\left(\begin{array}{cc} z_1 \qquad -{\overline z}_2 \\
z_2 \qquad {\overline z}_1 \end{array} \right)_i 
\label{ansatz2}
\ee
where $\chi_{\alpha\beta}$ are `particle-hole' matrix-valued 
operators~\cite{affleck}, 
$c_\alpha$, $\alpha=1,2$ are electron anihilation
operators, the Grassmann variables $\psi_i$, $i=1,2$ 
play the r\^ole of holon excitations, while the bosonic
fields $z_i, i=1,2,$ represent magnon excitations~\cite{Anderson}.
The ansatz (\ref{ansatz2}) 
has spin-electric-charge separation, since only the 
fields $\psi_i$ carry {\it electric} charge.
The
holon fields ${\widehat \psi} _{\alpha\beta}$ 
may be  
viewed as substructures of the 
physical electron $\chi_{\alpha\beta}$~\cite{quark},
in close analogy to the `quarks' of $QCD$.   
\pr
As argued in ref. \cite{fm} 
the ansatz is characterised by   
the following {\it local}
phase (gauge) symmetry structure: 
\be
  G=SU(2)\times U_S(1) \times U_E(1) 
\label{group}
\ee

The $U_E(1)$ electromagnetic symmetry is due to the electric
charge of the holons. In the absence 
of external electromagnetic potentials is a global symmetry (fermion number).
It becomes local (gauged) after coupling to external electromagnetism.

The 
local 
SU(2) symmetry is discovered if one defines the transformation 
properties of the ${\widehat z}_{\alpha\beta}$ 
and ${\widehat \psi} ^\dagger_{\alpha\beta}$ 
fields to be given by left multiplication
with the $SU(2)$ matrices, and pertains to the spin degrees of freedom.

The 
local $U_S (1)$ `statistical' phase symmetry
allows fractional statistics of the spin and charge 
excitations. 
This is an exclusive feature
of the three dimensional geometry, and is similar in spirit
to the bosonization technique of the spin-charge 
separation ansatz of ref. \cite{marchetti}.
The presence of $U_S(1)$ 
allows the alternative possibility 
of representing the holes as slave bosons and  
the spin excitations as fermions.

In the model of \cite{fm}, this $U_S(1)$ is assumed 
strongly coupled, capable of holon $\widehat \psi$ pairing 
and (parity-preserving) mass-gap generation. 
The mass generation breaks 
chiral symmetry, which can be defined 
in three-dimensional theories with {\it even} number 
of fermion species~\cite{app}, 
as is the case of the model
of ref. \cite{fm}. 
However, as discusssed 
in \cite{RK,dor,fm}, this mass gap is not accompanied by 
any phase-coherence, given that the symmetry breaking is 
realized in the Kosterlitz-Thouless mode~\cite{KT}.

At this stage 
we would like to make 
an important comment,
concerning the {\it relativistic 
nature} of the effective model
discussed in \cite{fm}. 
{}From a condensed-matter view point,  
such relativistic systems 
would arise by a {\it linearization about specific points 
on the fermi surface} of the statistical system, such as nodes etc.
In this respect it is worthy of mentioning that 
recent experimental tests~\cite{tsuei} imply  
that the superconducting gap in the high-$T_C$ cuprates
is of $d$-wave type, with lines of nodes on the fermi surface. 
It is the linearization about such nodes, in the flux phase
for the $U_S(1)$ gauge field,  
that leads to Dirac spectrum for holon excitations, 
with the fermi velocity of holes playing the r\^ole 
of the limiting (`light') velocity, as suggested
above and in refs. \cite{dor,fm}. Then,  
as a result of the Kosterlitz-Thouless (KT) 
mechanism for superconductivity described in \cite{dor,fm},  
a fermion gap 
opens at those nodes, which, due to the absence of a local-order
parameter, respects the $d$-wave character of the 
superconducting state.
\pr
The pertinent long-wavelength lattice gauge model, describing the low-energy 
dynamics around such $d$-wave nodes, 
assumes the form~\cite{fm}: 
\bea
&~&H_{HF}=\sum_{<ij>} tr\left[(8/J)\Delta^\dagger_{ij}\Delta_{ji}
+ {\hat K}(-t_{ij}(1 + \sigma_3)+\Delta_{ij}){\widehat \psi}_j V_{ji}U_{ji}
{\widehat \psi}_i^\dagger\right] + \nn \\ 
&~&\sum_{<ij>}tr\left[ {\hat K}{\overline {\widehat z}}_iV_{ij}U_{ij}{\widehat z}_j\right] + h.c. 
\label{Hub}
\eea
where $J$ is the Heisenberg antiferromagnetic interaction, 
${\hat K}$ is a normalization constant, and 
$\Delta_{ij}$ is a Hubbard-Stratonovich field that linearizes
four-electron interaction terms in the original Hubbard model, 
and 
$U_{ij}$,$V_{ij}$ are the link variables for the $U_S(1)$ and 
$SU(2)$ groups respectively. 
The conventional lattice gauge theory form of the action (\ref{Hub}) 
is derived upon freezing the fluctuations of the $\Delta_{ij}$ 
field~\cite{fm}, and   
integrating out the (massive) magnon fields, $z$,
in the path integral. This latter operation yields 
appropriate Maxwell kinetic terms for the link variables 
$V_{ij}$, $U_{ij}$, 
in a low-energy derivative expansion
~\cite{IanA,Polyakov}.
On 
the lattice such kinetic terms 
are given by plaquette terms of the form~\cite{fm}:
\be   
\sum_{p} \left[\beta_{SU(2)}(1-Tr V_p) + \beta_{U_S(1)}(1-Tr U_p)\right]
\label{plaquette} 
\ee
where $p$ denotes sum over 
plaquettes of the lattice, 
and $\beta_{U_S(1)} \equiv \beta_1 $, $\beta_{SU(2)} \equiv \beta_2 = 
4\beta_1$ 
are the dimensionless (in units of the lattice spacing) 
inverse square couplings of the $U_S(1)$ and $SU(2)$ groups,  
respectively~\cite{fm}. The above relation between the $\beta_i$'s
is due to the specific form of the $z$-dependent terms in (\ref{Hub}),
which results in the same induced couplings $g_{SU(2)}^2=g_{U_S(1)}^2$.
Moreover,  there is a non-trivial connection of the 
gauge group couplings to ${\hat K}$~\cite{fm}: 
\be
{\hat K} \propto g_{SU(2)}^2 = g_{U_S(1)}^2 \sim J\eta
\label{connection}
\ee 
with $\eta = 1-\delta$, 
$\delta$ being the doping concentration in the sample~\cite{fm,dorstat}. 
To cast the 
symmetry structure in a 
form that is familiar to 
particle physicists, one may change representation 
of the $SU(2)$ group, and instead of working with $2 \times 2$ 
matrices in (\ref{ansatz2}), one may use a representation 
in which the fermionic matrices ${\widehat \psi}_{\alpha\beta}$ 
are represented as two-component (Dirac) spinors in `colour' space:
\be
{\tilde \Psi}_{1,i}^\dagger =\left(\psi_1~~-\psi_2^\dagger\right)_i,~~~~
{\tilde \Psi} _{2,i}^\dagger=\left(\psi_2~~\psi_1^\dagger
\right)_i,~~~~~i={\rm Lattice~site} 
\label{twospinors}
\ee
In this representation  
the two-component spinors ${\tilde \Psi} $ (\ref{twospinors}) 
will act as {\it Dirac spinors on the Lattice}, 
and the $\gamma$-matrix (space-time)
structure will be spanned by the irreducible 
$2 \times 2$ Dirac representation. 
By 
assuming a background $U_S(1)$ 
field of flux $\pi$ per lattice plaquette~\cite{dor},
and considering quantum fluctuations around this background
for the $U_S(1)$ gauge field, 
one can show that there is a Dirac-like structure 
in the fermion spectrum~\cite{Burk,AM,dor,dorstat}, 
which leads to a conventional 
Lattice gauge theory form for the effective low-energy Hamiltonian of the
large-$U$,  doped Hubbard model~\cite{fm}. 

In the above context, a strongly coupled
$U_S(1)$ group can dynamically generate a mass gap 
in the holon spectrum~\cite{fm}, which breaks the $SU(2)$ local symmetry
down to its Abelian subgroup $U(1)$ generated by the $\sigma_3$ matrix.  
{}From the view point of the statistical model (\ref{Hub}), 
the breaking of the $SU(2)$ symmetry down to its Abelian 
$\sigma_3$ subgroup may be interpreted as  
restricting the holon hopping effectively to 
a single sublattice, since 
the intrasublattice hopping is suppressed 
by the mass of the gauge bosons.  
In a low-energy
effective theory of the massless degrees of freedom 
this reproduces the 
results of ref. \cite{dor,Sha}, derived 
under a large-spin 
approximation for the antiferromagnet, $S \rightarrow \infty$, 
which is not necessary 
in the present approach.

The Kosterlitz-Thouless (KT) nature~\cite{KT} of the 
$U_S(1)$ induced 
mass gap (absence of 
local order parameter), is a characteristic feature of 
gauge theories in $2+1$ dimensions, as argued in \cite{RK,dor}.
When applied to our non-Abelian model~\cite{fm} it leads to unconventional
KT superconductivity, 
provided the gauge boson of the unbroken $U(1) \in SU(2)$ 
is massless. Due to the compactness of the $U(1)$ gauge group, however,
which is a distinctive feature of the non-Abelian gauge group 
nature of the spin-charge separation (\ref{ansatz2}), 
there are non-perturbative effects (instantons), which are responsible for giving the gauge boson 
$U(1)$ a small but finite mass~\cite{ahw}. This spoils superconductivity,
leaving only a phase, 
characterised by pairing among the 
holons, without the existence of phase coherence. It is one of the points 
of this article to argue that such a phase may provide a 
possible explanation of 
the so-called `pseudogap' phase of high-temperature superconductivity~\cite{underdoped}, an intermediate non-superconducting phase, lying between the 
antiferromagnetic and $d$-wave superconducting~\cite{tsuei} phases. 
A preliminary discussion on this issue appeared in ref. \cite{fkm2}. 

At this point, we would like 
to mention that other authors have also used 
relativistic
fermions to describe the underdoped or pseudo-gap 
phase of high-$T_c$ materials~\cite{marchettilu,fisher}.
Their approaches, however, are different from ours:
In ref. \cite{marchettilu}, 
relativistic charge excitations are used 
as in our model~\cite{fm}. Their relativistic nature 
is due to the 
adopted scenario 
that the fermi surface 
of the underdoped cuprates 
consists of four small pockets, centered around $(\pm \frac{\pi}{2}, 
\pm \frac{\pi}{2})$ in momentum space. 
However, 
the low-energy model used in that work, and the 
nature of the gauge symmetries involved, are different 
from our model. 
In the `nodal liquid' approach of ref. \cite{fisher}, on 
the other hand, 
the relativistic Dirac-fermion excitations around 
the four nodes of the putative
fermi surface in the 
underdoped situation  
are neutral, and, hence, 
from our point of view they correspond to spin degrees of freedom rather 
than holons. This leads to a different physical scenario 
for the pseudo-gap phase than 
the one discussed here and in ref. \cite{fkm2}.

In this article we shall discuss in some detail 
the phase structure of the $SU(2) \otimes U_S(1)$ gauge theory.
Despite the above motivation from condensed-matter physics, 
the analysis and the 
techniques used will be those of Particle Physics, thereby making the results
even applicable to Particle Physics applications of three-dimensional Gauge theories, such as Early Universe studies, or high-temperature field theories. 
In this respect we mention the work by Volovik~\cite{volovik}, which 
pursues the analogy between the Physics of superfluid Helium and that of the Early Universe,
in an attempt to suggest condensed-matter experiments that could 
shed light in the physics 
of an early stage of our Universe. We hope that our work in this 
article 
will serve the purpose of pointing out yet another 
condensed-matter example, that of high-temperature superconductors, 
which may be connected to Particle Physics. 

The structure of the article is as follows: in section 2 we review the basic symmety properties of the Lattice action, including a discussion on the issue of 
spontaneous and/or dynamical  
breaking of parity in the context of the applicability 
of the Vafa-Witten~\cite{Vafa} theorem on the lattice. 
In section 3 we derive part of the phase diagram 
of the $SU(2) \otimes U_S(1)$ model, 
in the strong-coupling regime of the $SU(2)$ gauge group. 
The analysis is a non-trivial application of standard 
Lattice strong-coupling expansions~\cite{kawamoto,sam}
to our model. 
In section 4 we review briefly the {\it unconventional} 
superconducting properties of the system 
upon coupling it to external electromagnetic fields.
Emphasis is placed on the Kosterlitz-Thouless type of breaking 
of the electromagnetic symmetry, which is not accompanied by 
phase coherence.
In section 5 we discuss briefly the r\^ole 
of instantons in destroying superconductivity, but maintaining
pairing and fermion gap formation.
In section 6
we discuss a possible application of the model  
to the physics of high-temperature 
superconductors, with emphasis 
on the abovementioned r\^ole of instantons in 
inducing a 
pseudo-gap phase. 
This is an exclusive feature 
of the non-Abelian model of \cite{fm}. 
The possibility 
of tuning the doping concentration
in the sample to reach 
supersymmetric points in coupling-constant space~\cite{diamand},
with interesting consequences,  
is also mentioned briefly.  
Moreover, the r\^ole of additional four-fermion interactions,
which may dominate the superconducting phase, is pointed out.
Conclusions and outlook are presented in section 7. 
Some technical aspects of our approach, such as rules of strong-coupling 
expansion, and the evaluation of a Jacobian in the transition from 
fermionic Lattice variables to mesonic fields, 
are given in three Appendices. 

\section{Basic Symmetry Structure of the $SU(2) \otimes U_S(1)$ theory} 

\subsection{The Lagrangian of the model and its symmetries} 

The  theory (\ref{Hub}) 
corresponds, after integrating out the magnon degrees of 
freedom $z$~\cite{fm}, 
to a (low-energy) lattice lagrangian given by~\cite{farak,mavfar}:
\bea
&~&S[{\overline \Psi},\Psi,V,U]  =\frac{K}{2} \sum_{i,\mu}[{\overline
\Psi}_i (-\gamma_\mu)  
U_{i,\mu}V_{i,\mu} \Psi_{i+\mu}  + {\overline \Psi}_{i+\mu}
(\gamma _\mu)U^\dagger_{i,\mu}V^\dagger_{i,\mu}
\Psi _i ] + \nn \\
&~& \beta _1 
\sum _{p} (1 - trU_p) + \beta _2 \sum _{p} (1-trV_p) 
\label{action}
\eea
where
$U_{i,\mu}=exp(i\theta _{i,\mu})$ 
represents the 
statistical $U_S(1)$ gauge field and 
$V_{i,\mu}=exp(i\sigma ^a B_a)$ is the $SU(2)$ gauge field.
The quantity 
$K \equiv {\hat K}|t_{ij}|$, with $|t_{ij}|=t$ assumed small~\cite{fm}. 
The fermions are 2 component spinors in {\it both} Dirac (Greek) and colour
(Latin) space, $\Psi \equiv \Psi^{\alpha}_{a}$, and the generators of the
$SU(2)$ group are the $2\times2$ Pauli matrices,
$\sigma_{i}^{ab}$, $i=1,2,3$. 
The Dirac matrices can also be taken to have the
Pauli matrix representation (we continue to write them as
$\gamma_{\mu}^{\alpha\beta}$, to distinguish them from the $SU(2)$
colour matrices). 
Here we have passed onto a three-dimensional 
Euclidean  lattice formalism, in which ${\overline \Psi}$ is identified 
with $\Psi ^\dagger$.
For completeness we mention that the (naive)
continuum lagrangian corresponding to (\ref{action})
is given by: 
\be  
{\cal L} = -\frac{1}{4}(F_{\mu\nu})^2 
-\frac{1}{4}({\cal G}_{\mu\nu})^2  +{\overline \Psi}D_\mu\gamma_\mu\Psi 
\label{su2action}
\ee

\noindent
where $D_\mu = \partial_\mu -ig_1a_\mu^S-ig_2\sigma^aB_{a,\mu}$,
and $F_{\mu\nu}$, ${\cal G}_{\mu\nu}$ represent the 
field strengths for the $U_S(1)$, $SU(2)$ gauge groups respectively.

\pr
There are two sets of bilinears which transform as triplets under a
$SU(2)$ transformation: 
\bea
&~&{\cal A}_1 \equiv -i[{\overline \Psi}_1 \Psi_2 
- {\overline \Psi}_2 \Psi_1]~, \quad 
{\cal A}_2 \equiv i[{\overline \Psi}_1 \Psi_2 
+ {\overline \Psi}_2 \Psi_1]~, \quad 
{\cal A}_3 \equiv {\overline \Psi}_1\Psi _1 
- {\overline \Psi}_2 \Psi_2~;  
\nn \\
&~&{\cal F}_{1\mu} \equiv {\overline \Psi}_1\sigma _\mu \Psi _2 +
{\overline \Psi}_2\sigma_\mu \Psi_1~, \quad    
{\cal F}_{2\mu} \equiv i[{\overline \Psi}_1\sigma _\mu \Psi _2 -
{\overline \Psi}_2\sigma_\mu \Psi_1]~, \nn \\
&~& {\cal F}_{3\mu} \equiv {\overline \Psi}_1\sigma _\mu \Psi_1
-{\overline \Psi}_2\sigma _\mu \Psi_2
\label{fbil}
\eea
\noindent
where ${\overline \Psi}_a \equiv \Psi_a^{\dagger\alpha}
\gamma_0^{\alpha\beta}$
and  two $SU(2)$ bilinear singlets given by, 

\be
{\cal A}_4 \equiv {\overline \Psi}_1\Psi_1
+ {\overline \Psi}_2\Psi_2, \qquad 
{\cal F}_{4,\mu} \equiv {\overline \Psi}_1\sigma _\mu \Psi_1
+ {\overline \Psi}_2\sigma _\mu \Psi_2~, \qquad \mu=0,1,2      
\label{singlets}
\ee

\pr
In this approach one can define {\it meson} states~\cite{farak,fm}:
\be
   M_{ab,\alpha\beta} =\Psi_{b,\beta}{\overline \Psi}_{a\alpha}
\label{meson}
\ee
where Latin letters $a,b$ denote colour $SU(2)$ indices, and Greek letters 
$\alpha,\beta$ denote Lorentz spinor indices. 
One can re-express the meson state M, which is a $2 \times 2$ matrix
in both colour and Dirac space, in terms of the above bilinears \cite{farak};
\bea
M= &{\cal A}_3{\bf 1}.\sigma_3
+ {\cal A}_1{\bf 1}.\sigma_1 + {\cal A}_2 {\bf
1}.\sigma_2 + 
{\cal A}_4 {\bf 1}.{\bf 1} + \nn \\
&{\cal F}_{3,\mu}\gamma^\mu \sigma_3 +{\cal F}_{1,\mu}\gamma^\mu.\sigma_1
+ {\cal F}_{2,\mu}\gamma^\mu\sigma_2 + 
 {\cal F}_{4,\mu}\gamma^\mu{\bf 1} 
\label{mesontripl}
\eea 
\noindent
The first matrix written is in Dirac space, and the second is in
`$SU(2)$ colour' space.

\begin{figure}[htb]
\begin{center}
\parbox[c]{3in}{\psfig{figure=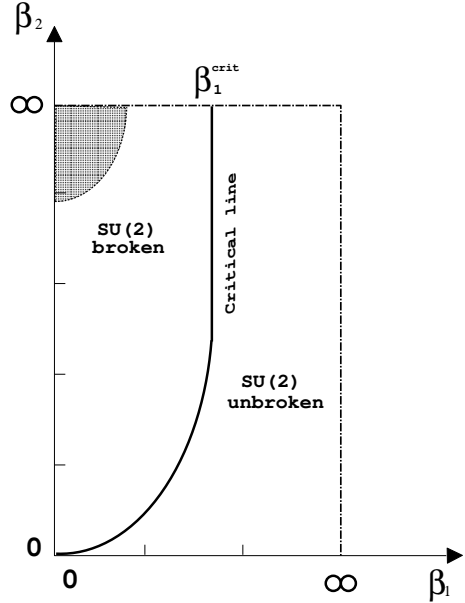,height=8cm,width=6cm}}
\end{center}
\caption{{\it Phase diagram for the $SU(2)\otimes U(1)$ 
model. The solid line is the critical line which 
is determined in this work, separating the 
phases of broken $SU(2)$ gauge symmetry from the phase where the 
symmetry is unbroken. 
Its precise shape is conjectural at present. Analytical 
and continuity arguments 
in this work determine the shape of the line in the 
neighborhood of $(\beta_1,\beta_2)=(0,0)$ and $(\beta_1,\beta_2)=
(\beta_1^c,\infty )$ only. 
This critical line 
also seems to characterise a solid state model, whose low-energy 
continuum limit is the gauge theory studied in this work.}}
\label{pdiag}
\end{figure}

The interesting feature of the $SU(2)\times U_S(1)$ model is that 
the parity-invariant condensate transforms as 
a $SU(2)$ triplet (\ref{fbil}), and, hence, once 
formed, it breaks $SU(2) \rightarrow U(1)$ {\it dynamically}~\cite{farak,fm}. 
The parity-violating condensate, one the other hand, is an $SU(2)$ singlet. 
In continuum theories, the {\it energetically} preferable configuration
in the {\it absence of external sources} is the parity-invariant 
condensate, 
according to the theorem of Vafa and Witten~\cite{Vafa,app} 
on the impossibility of spontaneous parity 
breaking in {\it vector like} theories, which we shall discuss 
in the next subsection. 
Thus, at least from naive continuous considerations, one expects that 
{\it Energetics} favours the formation of 
parity invariant condenates, 
and this was the main reason 
why parity violating condensates have been ignored, so far, 
in the existing literature. As we shall argue in the next subsection,
this feature is respected by the {\it lattice} model of \cite{fm}.

All these ideas can be incorporated into a rough phase diagram for a 
three-dimensional $SU(2) \times U(1)$ theory,
proposed in \cite{mavfar}. The 
diagram is depicted in fig. \ref{pdiag}. 
The
couplings shown are {\it
inverse couplings}, $\beta_2=4/ag_2^2$, $\beta_1=1/ag_1^2$,
where $a$ is the lattice spacing.

The top line $\beta_2=\infty$, $\beta_1 \ne 0$, 
corresponds to the $QED_3$ case. 
For $QED_3$ it is now generally accepted that there 
exists~\cite{app,IanA,maris}  
a critical
number of fermion flavours, below which there is dynamical formation
of a chiral condensate and chiral symmetry breaking
\cite{app,kondo}. In the language of an effective theory, where the 
dimensionless coupling is taken to be the inverse of the number of fermion
flavours \cite{IanA}, we can say that there is a {\it critical
coupling} above which there is symmetry breaking.

In figure \ref{pdiag}, the (inverse) critical
coupling on the lattice is denoted as 
$\beta_1^c$. The shaded area shows the weakly coupled
$SU(2)$ breaking, and the fact that we have no breaking on the
$\beta_2=0$ (ie infinitely coupled) $SU(2)$ line, as discussed in 
\cite{mavfar} and will be reviewed below, 
means -by continuity- that 
one 
can draw a 
tentative critical line separating the broken and unbroken phases. The
issue of whether the point where the 
line hits the $\beta_2=0$ axis is at the origin or at a finite value
of $\beta_2^c$ is one which we shall resolve here.

We note at this point that, in 
the context of our statistical model~\cite{fm}, there is the
special relation (\ref{connection}) among the 
(inverse) couplings of the $SU(2)$ and $U_S(1)$ factors, namely 
$\beta_2 = 4 \beta _1$, which, as we have mentioned, 
orginates from the special structure of the magnon ($CP^1$) 
degrees of freedom of the model. This special relation is 
interesting in that, when combined with the fact that 
the gauge couplings in the statistical model depend on the doping concentration
of the superconducting system, 
implies the existence of extreme values for the doping concentration,
above or below which the broken-$SU(2)$ gapped phase is lost. 
As we shall argue in this work, the critical value $\beta_2^{crit}=0$,
which implies that in the context 
of the present effective theory 
one cannot see a minimum coupling below which the $SU(2)$ symmetry is 
restored. 
It is understood that in the condensed matter context such a minimum coupling,
appropriate for the onset of antiferromagnetism, arises from the magnon 
$CP^1$ sector. We shall discuss such issues in more detail in section 6.

\subsection{Parity and Fermions on the Lattice}
\subsubsection{The Vafa-Witten Theorem in the Continuum}

Before embarking into a detailed analytical study of the
phase structure of $SU(2) \otimes U_S(1)$ theory,
we would like first to 
devote some time on the important issue of parity symmetry for {\it
Lattice } gauge models. 

As is well known, 
in continuum models, an important theorem, due to Vafa and Witten~\cite{Vafa}, 
forbids the spontaneous breaking of parity symmetry in vector-like theories,
in the sense that the parity-violating condensate
is not energetically preferable.  
Let us briefly review this, in the context of our three-dimensional gauge 
model~\cite{fm}.
We shall consider 
the Euclidian path integral for the two different mass terms,
corresponding to the condensates
$A_3$ (parity preserving), and $A_4$ (parity violating). 
Let us start from the case where the 
$<{\cal A}_4> \neq 0$ condensate is formed. 
In this case, the relevant path integral reads: 

\bea
&~&Z_{A_4}= \int DA D\overline{\Psi} D\Psi exp \int d^{3}x [L[A] +
\overline{\Psi}(i \! \not \!\!D + im) \Psi] = \nn \\
&~& \int DA \;det[i \! \not \!\!D + im] \; exp \int d^{4}x L[A]
\eea
where $L[A]$ denotes the pure gauge part of the lagrangian.

We can see that $det[i \! \not \!\!D]$ is positive because given $i \! \not
\!\!D \Psi = \lambda \Psi$ then $i \! \not
\!\!D (\gamma_2 \sigma_2 \Psi^*) = \lambda (\gamma_2 \sigma_2
\Psi^*)$. Thus every eigenvalue is repeated twice and the determinant
(the product of the e-values) is therefore real and positive. The gamma
matrices are in Dirac space and the sigma matrices are in colour
space.

However, $det[i \! \not \!\!D + im]$ is {\it not real} for the
following reason: The eigenvalue eiquations, in the presence of the mass $m$
read: 

\be
(i \! \not \!\!D + im) \Psi = (\lambda + im) \Psi ~, \;\;\;\;\;{\rm and}\;\;\;\;\; 
(i \! \not \!\!D + im) (\gamma_2 \sigma_2 \Psi^*) = (\lambda +im)
(\gamma_2 \sigma_2 \Psi^*)
\ee

The two equations have the same eigenvalue, however on
squaring each eigenvalue we get a complex number and therefore the
determinant is complex.

\pr
Let us now come to the case where the parity-invariant condensate is formed 
$ <{\cal A}_3> \neq 0$. In this case, the effective action reads: 

\be
Z_{A_3}= \int DA \;det[i \! \not \!\!D + im \sigma_3] \; exp \int d^{3}x S[A]
\ee
Applying the same method again we now get 
\be
(i \! \not \!\!D + im \sigma_3) \Psi = (\lambda + im) \Psi ~,
\;\;\;\;\; {\rm and} \;\;\;\;\;(i \! \not \!\!D + im \sigma_3) (\gamma_2
\sigma_2 \Psi^*)  =  (\lambda - im)(\gamma_2 \sigma_2 \Psi^*)
\ee
Now we see that the eigenvalues come in complex conjugate pairs, and
therefore the determinant is real and positive. 

Thus,  
the determinants in both cases have
the same absolute value, but the determinant in the case of a 
parity-violating condensate 
has an extra phase factor making
it complex. It is then straightforward to argue that the parity-violating 
case will not be energetically preferable~\cite{Vafa}. 
To this end, we note that 
the vacuum energy (in a box of volume $V$) is given by,

\be
e^{- E_j V} = \int d \mu (A) e^{\int d^3x L[A]} det_j  
\ee
where $det_j, j=3,4$ denotes the result of the fermion determinant 
in the case where the $A_3$ or $A_4$ condensates
are formed, respectively, and  
$d \mu (A)$ is the measure for the gauge field integration and is
positive; the same is true for 
the exponential $e^{\int d^3x L[A]}$ in the Euclidean formalism. 
By a generic result in complex integration calculus, then,  
the phase in $det_4$ can only make the integration
smaller than that for $det_3$, and therefore the vacuum energy associated with
mass $<{\cal A}_4>$ is larger than the vacuum energy for $<{\cal
A}_3>$. So we can say that the {\it energetically preferred} mass 
term is the parity
conserving one, and this is essentially the 
theorem of Vafa and Witten~\cite{Vafa}. Caution should be 
expressed in applying the 
theorem to the case of dynamical mass generation, due to the absence of 
bare fermion masses, which leads to the existence of fermion zero modes,
that make the Dirac operator ill defined, in need of regularization.
However, the rigorous analysis of ref. \cite{Vafa} deals with that case too.  

\subsubsection{Wilson fermions and the breakdown of the Vafa-Wittem 
theorem on the lattice}

On the lattice, however, 
the issue is non trivial, and still unsettled. 
As argued recently~\cite{parlat}, 
although the Vafa-Wiiten theorem~\cite{Vafa} may hold in the 
continuum limit,
however, on the Lattice there may be terms (at least in an effective Lagrangian level), proportional to the lattice spacing $a$, which may violate explicitly the parity symmetry, thereby acting like external sources
and hence spoiling basic assumptions of the Vafa-Wiiten theorem~\cite{Vafa}. 
The issue of how the continuum limit is taken is therefore a tricky one, and currently there is a debate as to whether spontaneous breaking of parity 
occurs on the lattice~\cite{parlat,bitar}.  
Although we shall not enter this debate, which concerns Wilsonian 
fermions on the lattice that we do not use here, however we consider it as useful to point out the difficulties associated with the parity symmetry, since 
it is a very important issue
for the 
superconductivity mechanism of the models~\cite{dor,fm}. 
This will help the reader appreciate better how these problems
are avoided in the specific lattice model of ref. \cite{fm}, which we use 
for our purposes in this work.  

The main problem with the lattice formulations of the Vafa-Witten theorem, 
using Wilson fermions, is associated with the fact that, in the case of 
spontaneous breaking of parity, 
the Dirac operator has zero modes, as we shall discuss below, and thus 
needs regularization. 
Such a regularization is provided by adding appropriate sources (which may trigger parity breaking) 
in the effective action and then removing them. 
The presence of a source term violates the vector-like nature of the regularized theory, and in general the problem arises 
from commuting the limits of removing the source or sending the bare mass term to zero. 

Let us first review the situation in the case of four-dimensional 
gauge theories. The reduction to three-dimensional gauge theories with 
even number of fermionic species will be straightforward, as we shall argue below. 
In the case of lattice regularization with Wilson fermions, 
the appropriate hermitean operator is not the Dirac operator but the 
overlap Hamiltonian $\gamma _5 W(m_0)$, where $m_0$ is a bare mass term 
needed for regularization of the Wilson-Dirac  operator $W$. 
This operator is 
known to have fermionic zero modes. The latter lead to a non-zero 
spectral density 
of eigenvalues
$\rho (\lambda, m_0)$ around $\lambda =0$, in the limit 
of zero (bare) fermion mass. 
If ${\tilde \rho}$ denotes 
the density of the Dirac operator $-i\nd{D}$,  then 
the following result holds~\cite{banks}:
\bea
     \rho (\lambda, m)&=& \frac{|\lambda|}{\sqrt{\lambda ^2 - m^2}}{\tilde 
\rho }(\sqrt{\lambda ^2 - m^2})~~~~|\lambda | >m \nn \\
&=& 0 ~~~~~~~ |\lambda | \le m 
\label{spectral}
\eea
As the fermion mass $m$ tends to zero, 
the operator $\rho (\lambda, m) \rightarrow {\tilde \rho}(\lambda) $
{\it non uniformly}.

To trigger numerically spontaneous breaking of a symmetry one adds
an appropriate symmetry-breaking source term and then removes it. 
In the case at hand, one should add a source term of the form 
$ih {\overline \Psi} \gamma _5 \sigma _3 \Psi $.  
As noted in \cite{bitar}, then,  
the parity-violating condensate, proposed to occur in Wilson fermions~\cite{parlat}, is proportional to $\rho (0,m_0)$ as the source term 
is removed, 
$h \rightarrow 0^{\pm}$: 
\be 
     < i {\overline \Psi} \gamma _5 \sigma _3 \Psi > =\mp 2\pi \rho (0, m_0)
\label{aoki}
\ee
where $m_0$ is a bare mass term which should be removed, and for simplicity 
we assumed two fermion flavours. 

The debate in the current literature~\cite{parlat,bitar} 
concerns the ordering of the limits $m_0 \rightarrow 0, 
h \rightarrow 0$ in the case of Lattice theories with Wilson fermions. The presence of a non-zero physical mass $m_0$ 
renders the limit $h \rightarrow 0$ safe, 
the parity-violating condensate vanishes on {\it both} 
the lattice and the continuum formalisms, 
and, thus, there is no problem 
with the theorem of \cite{Vafa}. 

In three-dimensional lattice gauge theories, 
with an {\it even} number of 
fermionic species, as the models we are interested in, chiral 
and parity-symmetry breaking may be studied in 
full analogy with four-dimensional gauge 
theories, provided one works with
a $4 \times 4 $ reducible Dirac representation~\cite{app},
generated by the following $\gamma $ matrices: 
\be
\gamma ^0=\left(\begin{array}{cc} {\bf \sigma}_3 
\qquad {\bf 0} \\{\bf 0} \qquad -{\bf \sigma}_3 \end{array} \right)~,
\qquad \gamma ^1 = \left(\begin{array}{cc} i{\bf \sigma}_1 
\qquad {\bf 0} \\{\bf 0} \qquad -i{\bf \sigma}_1 \end{array} \right)~, 
\gamma ^2 = \left(\begin{array}{cc} i{\bf \sigma}_2 
\qquad {\bf 0} \\{\bf 0} \qquad -i{\bf \sigma}_2 \end{array} \right)
\label{4gamma}
\ee
In such a case, there are two matrices that anticommute with the 
set of the $\gamma$-Dirac matrices~\cite{app}, 
\be
\gamma _3 \equiv  \left(\begin{array}{cc} 0 \qquad {\bf 1} \\
{\bf 1} \qquad 0 \end{array}\right),~  
\gamma _5 \equiv i\left(\begin{array}{cc} 0 \qquad {\bf 1} \\
{\bf -1} \qquad 0 \end{array}\right),~,
\label{gamm5}
\ee
Chiral symmetry is then generated by $\gamma_5$, 
and is broken by the parity-invariant condensate, which in four-component 
notation for the spinors $\Psi$ reads:
$A_3=<{\overline \Psi} \Psi >$. On the other hand, 
the 
parity-violating 
fermion condensate is given by: $A_4 =<{\overline \Psi} \Delta \Psi>$,
with: 
\be 
\Delta  \equiv i\gamma_3\gamma_5 =
\left(\begin{array}{cc} {\bf 1} \qquad 0 \\
0 \qquad {\bf -1} \end{array}\right) 
\label{gammamatr2}
\ee
It is, then, straightforward, following 
\cite{bitar}, to show that $A_4$ obeys a relation 
of the form (\ref{aoki}) for the three-dimensional case. 

We now remark that, 
in our condensed-matter inspired case (\ref{Hub}), the fermions describing 
holon nodal excitations in the $d$-wave state are {\it Dirac Spinors}
(\ref{twospinors}) and {\it not Wilson}. According to the discussion in \cite{aoki},
for such a case there are no consistency problems or 
ambiguities, as far as  results in the continuum 
are concerned~\cite{Vafa}. 
To show this, in our case, one should first note that, 
as we shall discuss in more detail in the next section,  
the effective potential for the meson fields $M$ (\ref{meson}) 
assumes the following generic form~\cite{farak,mavfar}:
\be 
V_{eff} \sim {\rm Tr} \sum_{i} \left( A {\rm ln}M_i   
- \sum_{\mu} P(M_i(-\gamma _\mu) M_i^\dagger \gamma _\mu)\right) 
\label{effpotent}
\ee
where $A$ is a numerical constant, 
$P (x)$ denotes an appropriate polynomial in $x$, $i$ is the 
lattice site, and ${\rm Tr}$ is taken over the (reducible) $4\times 4$ 
Dirac indices. As mentioned, above, to study spontaneous  parity breaking
numerically, one should add to (\ref{effpotent})  
an appropriate source term:
\be
 V_S \equiv hM_i 
\label{source}
\ee
Following ref. \cite{aoki}, we assume the following form 
for the vacuum wave function of 
$M_i$:
\be
     M_i = U e^{i \theta \Delta} = U\left( {\rm cos}\theta +  
i \Delta {\rm sin} \theta  \right) 
\label{maoki}
\ee
where $\Delta$ has been defined in (\ref{gammamatr2}).  

{}From the specific $M$-dependence of $P(x)$ in (\ref{effpotent}),
one observes that the only dependence on $\theta$ comes through 
the source term: 
\be
   V_{eff}[h] \sim 4\left( hU{\rm cos}\theta + A{\rm ln}U - 3P(U^2) \right)
\label{effp2}
\ee 
By extremizing (\ref{effp2}) with respect to $\theta$,
one obtain only the trivial minimum $\theta =0$, in constrast to the case 
of Wilson fermions, where the possibility 
for a non-trivial 
solution for $\theta$ exists, due to the 
Wilson parity-breaking term~\cite{aoki}. This solution 
implies that parity cannot br broken spontaneously, in agreement 
with the (continuum) theorem of Vafa and Witten~\cite{Vafa}. 

Hence, for our purposes in this work 
from now on we assume that {\it dynamical mass generation} in our 
$SU(2) \times U_S(1)$ 
model selects -due to energetics - the parity-invariant combination, 
which is accompanied by the presence of Goldstone Bosons due to 
the breaking of $SU(2) \rightarrow U(1)$~\cite{farak,fm,mavfar,fk}. 

\section{Study of the Phase Diagram of the 3-d $SU(2) \otimes U_S(1)$
Lattice Gauge Theory}

\subsection{The Phase Diagram in the regime $\beta_2=\infty$, $\beta_1$ 
arbitrary}

In this regime of the 
phase diagram the system is
equivalent to a strongly coupled $QED_3$, with global $SU(2)$
symmetry. This global symmetry acts like a flavour symmetry, and
is represented by an index $f$. Because of this limiting
procedure to $QED_3$, 
the spinors will be kept two component (\ref{twospinors}),
so that a smooth transition to the 
case of non-zero $SU(2)$ coupling is guaranteed. 
The pertinent path integral is: 
\be
Z = \prod_{i,\mu}\!\int dU_{i,\mu} d{\overline \Psi}^f_{i}d \Psi^f_{i} exp(-S[{\overline \Psi},\Psi,U]),
\label{}
\ee
where $S$ denotes the action (\ref{action}), written as

\bea
S[{\overline \Psi},\Psi,U] &=&\frac{1}{2}K \sum_{i,\mu}\sum_f[
\overline{\Psi}_{i}^{f}(-\gamma_{\mu}) \Psi_{i+\mu}^{f}U_{i,\mu} +
\overline{\Psi}_{i+\mu}^{f} \gamma_{\mu} \Psi_{i}^f U^\dagger_{i,\mu}]
\nn \\ 
&& + \beta _1 
\sum _{p} (1 - trU_p) \;\;.
\label{act2}
\eea

\pr
Putting $\beta_1=0$, we can do the $U_{S}(1)$ integral immediately
\cite{farak};

\bea &&
\prod_{i,\mu} \prod_f \!\int dU_{i,\mu} exp[
-\frac{K}{2} \sum_{i,\mu} (\overline{\Psi}_{i}^{f}(-\gamma_{\mu})
\Psi_{i+\mu}^{f} U_{i,\mu} +
\overline{\Psi}_{i+\mu}^{f} \gamma_{\mu} \Psi_{i}^f
U^\dagger_{i,\mu})], \nn \\
&=& \prod_{i,\mu} \prod_f I_{0}^{tr}(2\sqrt{y_{i,\mu}^f})\;\;.
\eea

\bea
y_{i,\mu}^f &\equiv& \frac{K^2}{4} \overline{\Psi}_{i}^{f}(-\gamma_{\mu})
\Psi_{i+\mu}^{f} \overline{\Psi}_{i+\mu}^{f} \gamma_{\mu} \Psi_{i}^f,
\nn \\
&=& \frac{K^2}{4} Tr[M_i^f \gamma_{\mu} M_{i+\mu} \gamma_{\mu}]\;\;.
\label{action4}
\eea
The quantity $(M_i^f)^{\alpha\beta}$ denotes the meson field (making the
Dirac indices explicit) $(\Psi_i^f)^{\alpha}
(\overline{\Psi}_i^f)^{\beta}$.
$I^{tr}_0$ denotes 
the zeroth order Modified Bessel function~\cite{Abra}, 
{\it truncated }
to an order determined by the number of the Grassmann (fermionic) 
degrees of freedom in the problem~\cite{fm,mavfar}.
In our case, due to the Dirac indices and 2 flavours of the {\it 
lattice spinors} $\Psi$, one should retain terms in $I^{tr}_0$  
up to ${\cal O}(y^4)$:

\be
I^{tr}_0(2\sqrt{y_{i\mu}})=
1+y_{i\mu}+\frac{y^2_{i\mu}}{4}
+\frac{y_{i\mu}^3}{36} +\frac{y_{i\mu}^4}{576} ,
\label{trunclog2}
\ee

The path integral is flavour symmetric $Z=\prod_f Z_f$ and, hence, we may 
factor
out and ignore this dependence.
We wish to obtain a path integral for the meson field $M_i^f$,
which necessitates the 
inclusion of the Jacobian for the pertinent field transformation.
This is calculated in 
Appendix 2, following 
\cite{kawamoto}. The result for the partition function reads:

\bea
Z_f &=& \prod_{i,\mu} \int dM_i^f exp[-\sum_i \log det M_i^f +
\sum_{i,\mu} \log I^{tr}_0(2\sqrt{y_{i,\mu}^f})]  \nn \\
&=& \prod_{i,\mu} \int dM_i^f exp[\sum_{i,\mu} -\frac{1}{6}\log det
M_i^fM_{i+\mu}^f + \log I^{tr}_0(2\sqrt{y_{i,\mu}^f})]\;\;.
\eea
The effective potential 
depends on the variable $y$ defined in 
(\ref{action4}). To determine its form, 
in terms of the 
condensate, one writes the VEV, $<M_i^f> = m^fW$ where $W$ is a unitary
matrix. So $Det M_i^f \sim m^2$ and $y_{i,\mu} \sim K^2m^2/2$.

\begin{figure}[htb]
\begin{center}
\parbox[c]{4in}{\psfig{figure=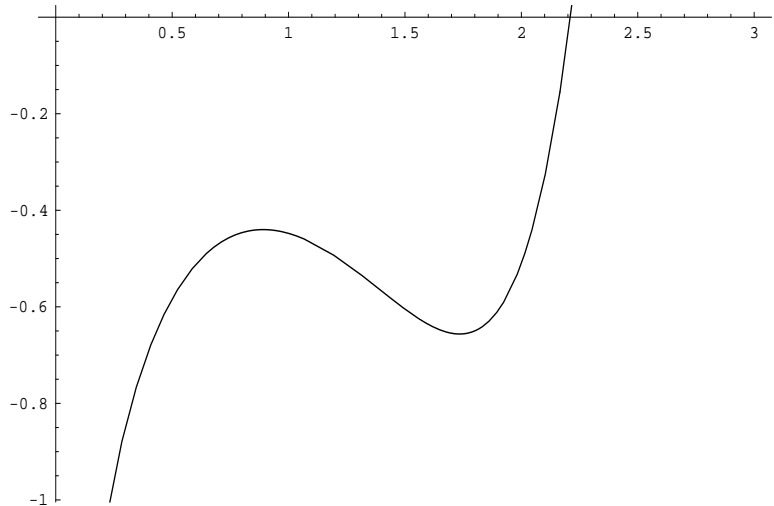,height=2.7in,width=4in}}
\end{center}
\caption{{\it The 
effective potential in the case $\beta_2=\infty$, $\beta_1=0$,
which coincides with the case of strongly coupled $QED_3$.
The potential has stationary points, implying 
a non-vanishing chiral-symmetry breaking condensate.}}
\label{figtwo}
\end{figure}

{}From the discussion in the previous section, we know that the 
energetically preferable configuration is the parity-conserving 
one, in which 
half of the fermion `flavours' acquire masses $+m$, and the rest 
acquire masses $-m$~\cite{Vafa,app}. Thus, 
the tree-level effective potential becomes, neglecting overall
factors, 
\be 
V_{eff} \sim \frac{2}{3}\log m - \frac{1}{2} K^2m^2 +
\frac{1}{16}K^4m^4 - \frac{1}{72}K^6m^6  
+ \frac{11}{3072}K^8m^8 ,
\label{smally}
\ee
which is plotted  in fig. \ref{figtwo} (with $K=1$)~\footnote{The 
general case $K \ne 1$ amounts to adding an irrelevant 
$K$-dependent constant to the 
expression for the effective potential. This is an exclusive feature 
of the minimal gauge model (\ref{action}), because in that case one 
may absorb
$K$ in the normalization of the fermion fields.  
However, in the presence of {\it additional} 
non-miminal fermion interactions, e.g. four-fermion interactions, which, as 
we shall see, 
may characterize
realistic models of doped antiferromagnets 
in their superconducting phases, the constant $K$ 
can no longer be absorbed in a normalization of the fermion fields, and hence
its magnitude acquires physical significance. We shall address such issues in section 6. For the purposes of this section, the 
minimal gauge model (\ref{action}) will suffice.}.  

This potential has stationary points and, from the arguments of
\cite{kawamoto} (described 
in Appendix 2), this is
sufficient to show that $<M_i^f> \neq 0$.

The incorporation of the gauge interactions 
will change the situation, and induce
non-trivial dynamics which may result in a change in symmetry
for some regions of the gauge coupling constants. 
{}From the previous result (\ref{smally}), and 
the discussion in section 2, it becomes clear 
that for weak $SU(2)$ and strong enough 
$U_S(1)$
the $SU(2)$ symmetry is broken down to a $U(1)$ subgroup~\cite{farak}. 
The non-trivial issue here 
is whether there exist critical (inverse) couplings $\beta_i^c, i=1,2$, 
above which  the symmetry is restored. 
According to earlier analyses,
either in the continuum or on the lattice~\cite{app,maris,kocic},
there appears to be 
a {\it critical coupling} $\beta_1^c$ 
on the axis $\beta_2=\infty$, $\beta_1={\rm free}$, 
above which dynamical mass generation
due to the $U_S(1)$ group cannot take place. This is depicted in 
figure \ref{pdiag}. 

In the large-$N$ continuum theories~\cite{app}
one usually identifies $8g_1^2 N = \alpha$, where $\alpha$ is kept fixed 
as $N \rightarrow \infty$, and plays the r\^ole of an effective 
ultraviolate cut-off. In our lattice action 
we may identify the inverse of our lattice spacing
$a^{-1}$ with $\alpha/8$, in which case $\beta_1^c \sim N_c$, 
with $N_c$ the critical number of (four-component) 
fermion flavours,
below which dynamical mass generation due to $U_S(1)$ occurs.  
To leading $1/N$ resummation~\cite{app}, 
$N_c \sim 32/\pi^2$; incorporating $1/N^2$ corrections 
shifts this number slighly higher. The issue on the 
existence of a critical number is still not quite settled, and  
proper 
lattice simulations are needed in this respect.
For our qualitative purposes, however, the large-$N$ critical number 
will be sufficient.

\subsection{Strongly-Coupled $SU(2) \otimes U_S(1)$ regime}

In this subsection we commence our analysis  
of the effects of strongly-coupled 
$SU(2)$ gauge interactions, by means of a small $\beta_2$ 
expansion.

\subsubsection{The Effective Action for $\beta_{2} = \beta_1 = 0$.}

First, we 
examine the model at the (limiting) 
point $\beta_{2} = \beta_1 = 0$ (ie the origin
of the diagram of Fig. \ref{pdiag}). 
We absorb the paramerter $K$ in a redefinition of the fermion fields,
because the action under consideration is only quadratic in $\Psi$
fields~\footnote{This is because our lattice fermions are of 
Dirac type. In the Wilson fermion case, on the other hand, 
due to the Wilson and bare mass terms, that one is forced to add in the 
lattice action, the above normalization of the spinors by $K$ cannot
be done.}. 
In this case, 
one may integrate out first the $U_S(1)$ gauge field. The $SU(2)$ 
action, then, is separable into an integral on each link on the 
lattice~\cite{mavfar}:  
\be
\int dV \left\{1 + tr({\overline A} V)tr(A V^{\dagger}) + \frac{
[tr({\overline A} V)tr(A V^{\dagger})]^2}{4} + \frac{
[tr({\overline A} V)tr(A V^{\dagger})]^3}{36} + \frac{
[tr({\overline A} V)tr(A V^{\dagger})]^4}{576} \right\},
\ee
where the variable $A$ are defined as follows:
\be 
A_\mu(x)^a_b = {\overline \Psi}_b(x+a) \gamma_\mu \Psi ^a (x), 
\qquad  {\overline A}_\mu(x)^a_b = 
{\overline \Psi}_b(x) (-\gamma_\mu)\Psi ^a (x+ a) \;\;.
\label{AAbar1}
\ee
The evaluation of these terms was first done by Samuel \cite{sam},
whose formalism we follow here. 
The resulting partition function, now with all gauge fields
integrated out, is:

\bea
Z_0= \prod_{i}\! \int d{\overline \Psi}_{i}d \Psi_{i} \prod_{i,\mu}\! 
& \left\{ 1+\frac{1}{2}tr(\overline{A}_{i,\mu}A_{i,\mu})+
\frac{1}{6}[tr(\overline{A}_{i,\mu}A_{i,\mu})]^2 
-\frac{1}{12}tr[(\overline{A}_{i,\mu}A_{i,\mu})^2]+  \right.\nn \\ 
& \left. + \frac{1}{48}[tr(\overline{A}_{i,\mu}A_{i,\mu})]^3
- \frac{1}{72}tr[(\overline{A}_{i,\mu}A_{i,\mu})^3]
+ \frac{17}{5760} [tr(\overline{A}_{i,\mu}A_{i,\mu})]^4 \right. \nn \\
& \left. - \frac{1}{320}
[tr(\overline{A}_{i,\mu}A_{i,\mu})]^2tr[(\overline{A}_{i,\mu}A_{i,\mu})^2]
+ \frac{1}{1920} tr[(\overline{A}_{i,\mu}A_{i,\mu})^2]^2 \right\}\;\;.
\label{zeroth}
\eea
Since we want $Z_0$ to be in the form $Z_0=e^{S_eff}$ we exponentiate the
above polynomial still keeping terms up order $O(A\overline{A})^4$. 

\bea
Z_0=\prod_{i}\! \int d{\overline \Psi}_{i}d \Psi_{i} \exp \sum_{i,\mu}&&
\left\{ \frac{1}{2}tr(\overline{A}_{i,\mu}A_{i,\mu})+
\frac{1}{24}[tr(\overline{A}_{i,\mu}A_{i,\mu})]^2
-\frac{1}{12}tr[(\overline{A}_{i,\mu}A_{i,\mu})^2] \right. \nn \\
&& -\frac{1}{48}[tr(\overline{A}_{i,\mu}A_{i,\mu})]^3  
- \frac{1}{72}tr[(\overline{A}_{i,\mu}A_{i,\mu})^3] \nn \\
&& 
+\frac{1}{24}
tr(\overline{A}_{i,\mu}A_{i,\mu})
tr[(\overline{A}_{i,\mu}A_{i,\mu})^2] + \frac{3}{640}
[tr(\overline{A}_{i,\mu}A_{i,\mu})]^4  \nn \\
&& - \frac{29}{2880}
[tr(\overline{A}_{i,\mu}A_{i,\mu})]^2tr[(\overline{A}_{i,\mu}A_{i,\mu})^2]
+\frac{1}{144}
tr(\overline{A}_{i,\mu}A_{i,\mu})[tr(\overline{A}_{i,\mu}A_{i,\mu})]^3
\nn \\ 
&& \left. -\frac{17}{5760}tr[(\overline{A}_{i,\mu}A_{i,\mu})^2]^2
\right\} \;\;.
\label{expact}
\eea
We want to rearrange the $({\overline \Psi},\Psi)$ which make up the $
\overline{A}, A $ to get an effective action in terms of meson fields
defined on site.
The standard procedure for making an effective action in terms of
meson fields is to rewrite each term in (\ref{zeroth}) as
a function of $M^{aa,\alpha\beta} = \Psi^{a, \alpha} {\overline
\Psi}^{a, \beta}$ eg;

\bea
tr(\overline{A}_{i,\mu}A_{i,\mu})& = &{\overline \Psi}_i ^{a, \alpha}
(-\gamma_{\mu})^{\alpha\beta} \Psi_{i+\mu}^{ b, \beta}
{\overline \Psi}_{i+\mu}^{b, \gamma} (\gamma_{\mu})^{\gamma\delta} 
\Psi_{i}^{a, \delta}, \nn \\
&=& -\Psi_{i}^{a, \delta} {\overline \Psi}_{i}^{a, \alpha} \Psi_{i+\mu}^{
b, \beta} {\overline \Psi}_{i+\mu}^{b, \gamma}
(-\gamma_{\mu})^{\alpha\beta} (\gamma_{\mu})^{\gamma\delta}, \nn \\
&=& - tr_c M_{i}^{\delta\alpha} tr_c M_{i+\mu}^{\beta\gamma}
(-\gamma_{\mu})^{\alpha\beta} (\gamma_{\mu})^{\gamma\delta}\;\;. 
\eea
$tr_c M$ means we have traced over the colour indices. However if we
were to substitute in for M the expansion in terms of bilinears
(\ref{mesontripl}) we would find that the term containing ${\cal A}_3$
vanishes because of the traceless property of the Pauli matrix $\sigma_3$.
Thus the above rearrangement of $tr(\overline{A}A)$ will not be of any use
if we want to get information on the VEV of ${\cal A}_3$. We must thus
look for an alternative rearrangement.

This alternative arrangement can be demonstrated if we look at the
term $[tr(\overline{A}_{i,\mu}A_{i,\mu})]^2$, 
which is written out as;

\bea
&~&[tr(\overline{A}_{i,\mu}A_{i,\mu})]^2= 
\nn \\
&~&[\overline{\Psi}_{i}^{b,\alpha}
(-\gamma_\mu)^{\alpha\beta}\Psi_{i+\mu}^{a,\beta}
\overline{\Psi}_{i+\mu}^{a,\gamma}
(\gamma_\mu)^{\gamma\delta} \Psi_{i}^{b,\delta}]
\times
[\overline{\Psi}_{i}^{d,\epsilon}
(-\gamma_\mu)^{\epsilon\zeta}\Psi_{i+\mu}^{c,\zeta}
\overline{\Psi}_{i+\mu}^{c,\theta}
(\gamma_\mu)^{\theta\eta} \Psi_{i}^{d,\eta}],
\eea
we can rearrarange the $(\Psi,\overline{\Psi})$ which appear into {\it
pairs} of meson states defined on each site.
\bea
&\Psi_{i}^{b,\delta} \overline{\Psi}_{i}^{d,\epsilon} \Psi_{i}^{d,\eta}
\overline{\Psi}_{i}^{b,\alpha} \Psi_{i+\mu}^{a,\beta}
\overline{\Psi}_{i+\mu}^{c,\theta} \Psi_{i+\mu}^{c,\zeta}
\overline{\Psi}_{i+\mu}^{a,\gamma} \gamma_\mu^{\alpha\beta}
\gamma_\mu^{\gamma\delta}  \gamma_\mu^{\epsilon\zeta}
\gamma_\mu^{\theta\eta}, \nn \\
=& tr_c (M_{i}^{\delta\epsilon} M_{i}^{\eta\alpha}) tr_c
(M_{i+\mu}^{\beta\theta}M_{i+\mu}^{\zeta\gamma})  \gamma_\mu^{\alpha\beta}
\gamma_\mu^{\gamma\delta}  \gamma_\mu^{\epsilon\zeta}
\gamma_\mu^{\theta\eta}\;\;.
\eea
This time when we substitute in our bilinear expansion for
$M_i^{ab\alpha\beta}$ we do not lose the part depending on ${\cal
A}_3$. We are no longer tracing over a single Pauli matrix, but rather
over $tr(\sigma_3^2) = tr(1)$.
This procedure will be justified if we find that there exists an
energetically favorable non-zero VEV for ${\cal A}_3$. 

Also we see that the terms in the action which have an odd
number of $(\Psi,\overline{\Psi})$ pairs (ie. $O(A\overline{A})$ and
$O(A\overline{A})^3$) vanish in this procedure
since they cannot be arranged to solely produce {\it pairs} of mesons. 
There is a subtlety involved in calculating the $O(A\overline{A})^4$
term, which we describe in Appendix 1.

Having followed this procedure for each term in the action the next
step will be to substitute in for $M$, or more precisely for ${\cal
A}_3$, the VEV we are looking for. We also assume that the VEVs of all
other bilinears are zero \cite{farak,fk,mavfar}. Therefore
$<M_i^{ab,\alpha\beta}> = U_i \sigma_3^{ab} \delta^{\alpha\beta}$.
Following the above procedure for each term in the action (\ref{expact})
we have:

\bea
<O(A\overline{A})> &=& 0, \nn \\
<\frac{1}{24}[tr(\overline{A}_{i,\mu}A_{i,\mu})]^2> &=& \frac{4}{3}
U_i^2 U_{i+\mu}^2, \nn \\
<-\frac{1}{12}tr[(\overline{A}_{i,\mu}A_{i,\mu})^2]> &=& \frac{4}{3}
U_i^2 U_{i+\mu}^2, \nn \\ 
<O(A\overline{A})^3> &=& 0, \nn \\
<\frac{3}{640} [tr(\overline{A}_{i,\mu}A_{i,\mu})]^4> &=& \frac{3}{5}
U_i^4 U_{i+\mu}^4, \nn \\ 
<-\frac{29}{2880}
[tr(\overline{A}_{i,\mu}A_{i,\mu})]^2tr[(\overline{A}_{i,\mu}A_{i,\mu})^2]>
&=& \frac{116}{135} U_i^4 U_{i+\mu}^4, \nn \\ 
<\frac{1}{144}
tr(\overline{A}_{i,\mu}A_{i,\mu})[tr(\overline{A}_{i,\mu}A_{i,\mu})]^3>
&=& \frac{4}{9} U_i^4 U_{i+\mu}^4, \nn \\ 
<-\frac{17}{5760}tr[(\overline{A}_{i,\mu}A_{i,\mu})^2]^2> &=&
-\frac{17}{54} U_i^4 U_{i+\mu}^4\;\;.
\eea

\begin{figure}[htb]
\vspace{0.2in}
\begin{center}
\parbox[c]{3in}{\psfig{figure=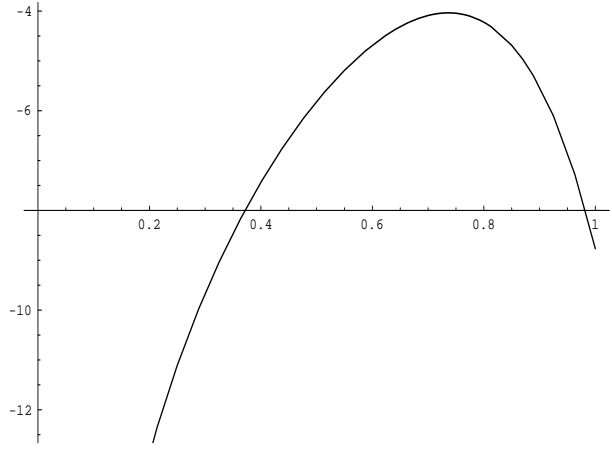,height=6cm,width=8cm}}
\label{effpot} 
\caption{{\it The effective potential
for $\beta_1=\beta_2=0$.}}
\label{figthree}
\end{center}
\end{figure}

One can now change path-integration variables, from fermion to meson fields
${\cal M}$. An important r\^ole in the dynamics of the system 
is played by the Jacobian of such a transformation, which is calculated in 
Appendix 2 following standard arguments~\cite{kawamoto}.  
Including the Jacobian we get
\bea
Z_0 &=& \prod_{i}\! \int d {\cal M}_i 
exp \left\{ -\sum_{i,\mu} (-\frac{4}{3} U_i^2U_{i+\mu}^2 -
\frac{143}{90} U_i^4 U_{i+\mu}^4)  - \sum_i 2\log U_i^4 \right\}, \nn \\
&=& \prod_{i}\! \int d {\cal M}_i 
exp \left\{ -\sum_{i,\mu} (-\frac{4}{3} U_i^2U_{i+\mu}^2 -
\frac{143}{90} U_i^4 U_{i+\mu}^4 + \frac{2}{6} \log U_i^4 U_{i+\mu}^4
)\right\},
\label{finaleff}
\eea
which describes the dynamics in the $\beta_2=\beta_1=0$ region of the
phase diagram of  
figure \ref{pdiag}. To get a simple saddle point effective potential
we complexify the condensate $U_i$ and 
write, following ref. \cite{kawamoto}, $U_iU_{i+\mu} = U^2$. This is
justified, since we assume that, in the general case, 
the condensate $U_i$ should be the same
for all odd sites and all even sites separately. In our path
integral the radial part of the contour is irrelevant (see Appendix
2) and we can choose $U_{even} = U e^{i\varphi}$ and $U_{odd} = U
e^{-i\varphi} $. As we discuss in Appendix 2, the minimum 
value of the effective potential occurs for $\varphi =0$. 
The zeroth order effective potential is defined $V_{eff} = -S_{eff}/vol$:

\be
V_{eff} = 8 ln U  - 4 U^4 - \frac{143}{30} U^8 ,
\label{zeroeff}
\ee
which is plotted in figure \ref{figthree}.

We observe that it has a local maximum, but, as explained in
\cite{kawamoto}, this still implies stability of the broken $SU(2)$
vacuum, due to the special properties of the Jacobian associated with
the transformation from the $\Psi, \overline{\Psi}$ variables to the
meson variables $\cal{M}$. This is reviewed briefly in Appendix 2.

{}From figure \ref{figthree} it is evident that there exists $SU(2)$
symmetry breaking even for the csae of $\beta_2 = 0$. This implies
that the critical line in the phase diagram of figure \ref{pdiag}
passes through the origin. This is also the situation argued to
characterize the statistical model of \cite{fm}, which may describe
high-temperature superconductivity.

\subsubsection{Strong Coupling Expansion of $\beta_{2}$}

We now look at the strong coupling expansion of the $SU(2)$ field,
up to and including order ${\cal O}(\beta_2^2)$, 

\be
exp[-\beta_2\sum_p(1-trV_p)]\simeq 1-\beta_2\sum_p(1-trV_p)+\frac{\beta^{2}_{2}}{2}\sum_p(1-trV_p)\sum_q(1-trV_q)+O(\beta_2^3),
\ee

\noindent
The zeroth order term has been calculated in the previous subsection
(\ref{finaleff}).  \\

\noindent
The first order term in the $SU(2)$ integral is written \\
\bea
&&Z_1 = -\beta_{2}\prod_{i,\mu}\!\int dV_{i,\mu}
I^{tr}_{0}(2\sqrt{y_{i,\mu}})(N_p -\sum_p trV_{p}), \nn \\
&=& -\beta_{2}N_p \prod_{i,\mu}\!\int dV_{i,\mu}
I^{tr}_{0}(2\sqrt{y_{i,\mu}}) =-\beta_{2}N_p Z_0,
\eea
because the addition of a single plaquette will give an odd number of
group elements on each side of the plaquette, and therefore will
integrate to zero. \\

\noindent
At the second order the calculation is no longer so simple,
\be
Z_{2}=\frac{\beta^{2}_{2}}{2}\prod_{i,\mu}\!\int dV_{i,\mu}
I^{tr}_{0}(2\sqrt{y_{i,\mu}}) \sum_p (1-trV_p) \sum_q (1-trV_q^\dagger),
\ee
by the same argument as used in the first order case, we can
only have a non-zero integral where we avoid integrals over different
numbers of $V$s and $V^{\dagger}$s. Thus the product
$\sum_p (1-trV_p) \sum_q (1-trV_q)$ can be replaced by
$(N_p^2-\sum_p trV_p^\dagger trV_p)$ 

\bea
Z_{2}&=&\frac{\beta^{2}_{2}}{2}\prod_{i,\mu}\!\int dV_{i,\mu}
I^{tr}_{0}(2\sqrt{y_{i,\mu}})(N_p^2 + \sum_p trV_{p}trV_{p}^\dagger),
\nn \\
&=&\frac{\beta_{2}^2N_p^2}{2}Z_{0}+\frac{\beta^{2}_{2}}{2}\prod_{i,\mu}\!
\int
dV_{i,\mu}I^{tr}_{0}(2\sqrt{y_{i,\mu}})\sum_ptrV_{p}trV_{p}^\dagger
\;\;. \nn \\
\eea

Hence, the problem is to calculate the group integrals which make up the
non-trivial part of $Z_2$ namely;

\bea
&& Z_2 = \prod_{i,\mu} \int
dV_{i,\mu} I^{tr}_{0}(2\sqrt{y_{i,\mu}})\sum_ptrV_{p}trV_{p}^\dagger \nn \\
&=& \sum_p \prod_{i,\mu \not\in p} \int
dV_{i,\mu} I^{tr}_{0}(2\sqrt{y_{i,\mu}}) \prod_{i,\mu \in p} \int
dV_{i,\mu} I^{tr}_{0}(2\sqrt{y_{i,\mu}}) trV_{p}trV_{p}^\dagger \nn \\
&=& \prod_{i,\mu} \int
dV_{i,\mu} I^{tr}_{0}(2\sqrt{y_{i,\mu}})
\sum_p \frac{\prod_{i,\mu \in p} \int
dV_{i,\mu} I^{tr}_{0}(2\sqrt{y_{i,\mu}}) trV_{p}trV_{p}^\dagger
}{\prod_{i,\mu \in p} \int 
dV_{i,\mu} I^{tr}_{0}(2\sqrt{y_{i,\mu}})}\;\; \nn \\
&=& Z_0 \sum_p J_p
\eea
The full path integral is then written as
\be
Z=Z_0(1-\beta_{2}N_p +
\frac{\beta_{2}^2N_p^2}{2})+\frac{\beta_{2}^2}{2} Z_0 \sum_p J_p\;\;.
\ee
We will be interested in the logarithm of this function:

\be
S_{eff} = \log \left[Z_0 \left\{1-\beta_{2}N_p +\frac{\beta_{2}^2N_p^2}{2}
+\frac{\beta_{2}^2}{2} \sum_p J_p \right\} \right],
\ee
which can be expanded, ignoring constant factors and keeping terms
up to $O(\beta^2)$

\be
S_{eff} = \log Z_0 + \frac{\beta^2_2}{2} \sum_p J_p \;\;.
\ee
Therefore, the strong coupling expansion leads us to calculate the group
integrals around the plaquette in the function $J_p$

\be
J_p =  \frac{\prod_{i,\mu \in p} \int
dV_{i,\mu} I^{tr}_{0}(2\sqrt{y_{i,\mu}}) trV_{p}trV_{p}^\dagger
}{\prod_{i,\mu \in p} \int 
dV_{i,\mu} I^{tr}_{0}(2\sqrt{y_{i,\mu}})}\;\;.
\ee
The denominator has already been given earlier and is the product
around each side of the plaquette of the $(A\overline{A})$ polynomial in
(\ref{zeroth}). For convenience let us label the 
plaquette as having sides 1,2,3,4 and sites A,B,C,D

\bea
\left\{\prod_{i,\mu \in p} \int 
dV_{i,\mu} I^{tr}_{0}(2\sqrt{y_{i,\mu}}) \right\}^{-1}& = &(1 +
\frac{1}{2}trA_{1}\overline{A}_{1} + \cdots)^{-1}
(1+\frac{1}{2}trA_{2}\overline{A}_{2} + \cdots)^{-1} \times \nn \\
&&
\times (1+\frac{1}{2}trA_{3}\overline{A}_{3} + \cdots)^{-1} 
(1+\frac{1}{2}trA_{4}\overline{A}_{4} + \cdots)^{-1},
\eea
These brackets can be expanded and truncated to $O(A\overline{A})^4$
since they still contain the grassmann fields $(\Psi\overline{\Psi})$.
The numerator contains a multitude of group integrals which need to
be evaluated around the plaquette. The method, outlined briefly in
Appendix 3,  is somewhat involved and
the interested 
reader can refer to \cite{danthes} for the algorithms used. Suffices to
say that the terms which have an even number of $(\Psi\overline{\Psi})$ are
kept since they will form the meson states.

After a tedious calculation along the above sketched lines, 
the quantity $J_p$ becomes

\bea
J_p =&1 + \frac{628}{81} (U_AU_BU_CU_D)^2 + \frac{1624}{135}
(U_AU_BU_CU_D)^2(U_A^2U_B^2 + U_B^2U_C^2 +U_C^2U_D^2 +U_D^2U_A^2) \nn
\\
& +
\frac{548069}{18225} (U_AU_BU_CU_D)^4,
\eea
and since in three dimensions we have one plaquette per link the effective
potential to $O(\beta^2_2)$ is (again with $U_iU_{i+\mu}=U^2$):

\be
V_{eff}(\beta_2^2) = 8 ln U  - 4 U^4 - \frac{143}{30} U^8 -\frac{\beta^2_2}{2}
\left\{ 3 + \frac{628}{27} U^8 + \frac{1624}{45} U^{12} +
\frac{548069}{18225} U^{16} \right\}\;\;.
\label{twoeff}
\ee

\begin{figure}[htb]
\vspace{0.2in}
\begin{center}
\parbox[c]{3in}{\psfig{figure=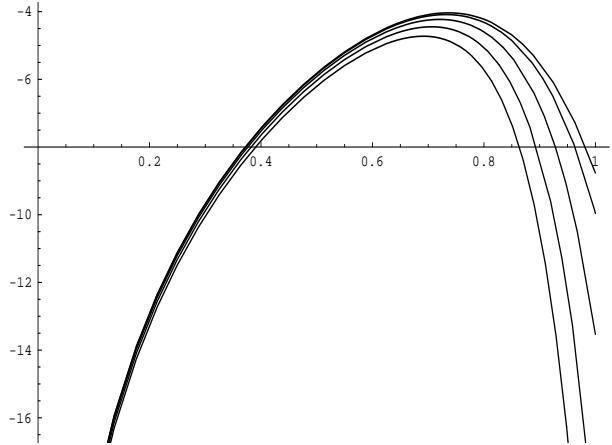,height=6cm,width=8cm}} 
\caption{{\it The effective potential
for $\beta_1=0$ and for $\beta_2=0,0.125,0.25,0.375,0.5$,
the corresponding curves lying in order of decreasing 
magnitude of their maxima 
from right to left in the figure.}}
\label{effpot2}
\end{center}
\end{figure}

This is plotted in fig. \ref{effpot2},
with $\beta_2$ taking a range of values between
$0$ and $0.5$. The behaviour does not change qualitatively as we
increase $\beta_2$, showing that the symmetry remains broken as we move
up the $\beta_2$ of figure \ref{pdiag}. This is as expected
assuming a continuous critical line.

\subsection{The Phase Diagram for $\beta_2 =0$, $\beta_1 \ne 0$}

Let us now complete our analysis on the phase diagram 
by concentrating on the region of strong 
$SU(2)$, $\beta_2 = 0$, keeping $U_S(1)$ coupling 
arbitrary (bottom horizontal axis of fig. 1). 
In this part of the phase diagram one 
can {\it integrate out} the (strongly coupled) $SU(2)$ gauge fields
to derive an effective action for the 
fermion and $U_S(1)$ gauge fields.
The $SU(2)$ path integration 
is performed along the lines of ref. \cite{kawamoto}.
In the strong coupling limit for $SU(2)$, $\beta_2 =0$,
the effective action,
obtained after integration 
of the $SU(2)$ gauge fields, reduces to the a sum of {\it 
one-link contributions}, $S_{eff}=S_{eff}(A,{\overline A})$,
with 
\be 
    A_\mu(x)^a_b = {\overline \Psi}_b(x+a) \gamma_\mu 
U^\dagger_{x,\mu}\Psi ^a (x) 
\qquad  {\overline A}_\mu(x)^a_b = 
{\overline \Psi}_b(x) (-\gamma_\mu) U_{x,\mu}\Psi ^a (x+ a) 
\label{AAbar}
\ee
where $U_{x,\mu}$ denotes the $U_S(1)$ group element, 
$a$ is the lattice spacing, and the latin indices $a,b$  
are colour $SU(2)$ indices. 
Below we shall proceed to evaluate explicitly this 
strongly-coupled 
effective action along the $\beta_2=0$ axis of 
the phase diagram in fig. \ref{pdiag}.

For the $SU(N)$ case 
the effective action $exp(-S_{eff})$ is known
in an expansion over 
$A$, ${\overline A}$~\cite{kawamoto}. This will be sufficient for our 
purposes here: 
\be
S_{eff}= \frac{1}{N}Tr({\overline A}A) 
+ \frac{1}{2N(N^2-1)}
[-Tr[({\overline A}A)^2] + \frac{1}{N}(Tr[{\overline A}A])^2]
+ \dots + \frac{1}{N!}(detA + det{\overline A}) + \dots 
\label{effAA}
\ee
The determinant terms are associated with baryonic states~\cite{kawamoto}.
We also note that 
for the $U(N)$ case the determinant terms are {\it absent}. 
In the phase diagram of fig. 1 the $U(2)$ case 
occurs at the point $\beta_2 \rightarrow 0, \beta_1 \rightarrow 0$. 
We approach this point 
asymptotically, by working on the $\beta_2=0$ line, and 
assuming $\beta_1 \ne 0$. 
We first notice that the Abelian phase factors of the $U_S(1)$ interactions
{\it cancel} from the expressions for the traces 
of $A$, ${\overline A}$ in the effective action (\ref{effAA}).
Moreover, from the discussion 
of section 2, we know that 
the $SU(2)$ (strong-coupling) integration  
{\it cannot produce} a parity-invariant 
condensate, since the latter is not an $SU(2)$ singlet~\cite{farak}. 
The resulting effective action should be expressible in terms 
of $SU(2)$ invariant fields. 
Thus, on the axis $\beta_2=0$ there is {\it no 
possibility} for the $U_S(1)$ group to generate a fermion condensate. 
This implies that for very strong $SU(2)$ group the
symmetry is {\it restored} for arbitrary $U_S(1)$ couplings. 

This is a very important fact, indicating the existence of a
not well-defined limit $\beta_2, \beta_1 \rightarrow 0$, since from the  
discussion in this and the previous subsections 
it seems that there is an ordering problem in how  
one approaches the point $(\beta_1, \beta_2)=(0,0)$. This indicates that
the shape of the critical line around that point 
is the one depicted in fig. \ref{pdiag},
concaving upwards. By {\it continuity arguments}, then, one expects
the shape of the entire critical line to be the one depicted in the figure. 
About the point $(\beta_1,\beta_2)=(\beta_1^c,0)$, where the
$SU(2)$ interactions are negligibly weak, and thus irrelevant, 
one expects an almost 
vertical shape of the critical line. 

This discussion completes our analytical results 
for the phase diagram of the 
$SU(2) \otimes U_S(1)$ gauge theory. 
As we have already mentioned above, the derivation of the precise 
shape of the 
critical line,
separating the phases of unbroken $SU(2)$ symmetry from the region where 
symmetry breaking occurs, requires a proper lattice simulation analysis, 
by means of a fermionic algorithm. We hope to 
be able to address these issues  
in a future publication. 
However, the 
above results 
will be sufficient for our purposes in this work,
and will enable us to present physically intreresting 
scenaria, pertinent to 
the physics of high-temperature superconductivity,
which we shall discuss in section 6.

Before doing so, it will be essential to review 
in the next two sections : (i)  
the 
symmetry breaking 
properties of our non-Abelian gauge model
from  
the point of view of Goldstone's theorem and 
the existence of a local order parameter,
and (ii) the r\^ole of 
{\it non perturbative effects}. 
These will be important in considering
the  
coupling 
of the model 
to external electromagnetic fields, as required 
for the study of superconducting properties.

\section{Kosterlitz-Thouless Realization of 
Superconductivity in the $SU(2) \otimes U_S(1)$ model}

This section is mainly a review of results that appear
in the literature regarding the model~\cite{fm,dor,RK}. 
It mainly serves as a comprehensive account,
for the benefit of the non-expert in the area,  
of the various delicate
issues involved, which play a very crucial r\^ole in the underlying 
physics. 

An important issue in the model (\ref{su2action}) is the 
existence of a {\it global conserved symmetry}, namely the fermion 
number, which is due 
to the electric charge of the fermions $\Psi$.
The corresponding current is given by 
\be
J_\mu = \sum_{c=1}^{2} {\overline \Psi}^c \gamma _\mu \Psi _c
\label{fermionnumber}
\ee
This current generates a global $U_E(1)$ symmetry, 
which 
after 
coupling with external electromagnetic fields 
is {\it gauged}. 

In the absence of such external potentials, the 
symmetry $U_E(1)$ 
is {\it broken spontaneously} 
in the massive phase for the fermions $\Psi$.
This 
can be readily 
seen by considering the following matrix element (see figure \ref{ktbreak}):

\begin{centering}
\begin{figure}[htb]
\vspace{2cm}
%
\bigphotons
\begin{picture}(30000,5000)(0,0)
\put(20000,0){\circle{100000}}
\drawline\photon[\E\REG](22000,0)[4]
\put(18000,0){\circle*{1000}}
\end{picture}
%
\vspace{1cm}
\caption{{\it Anomalous one-loop Feynman matrix element,
leading to a Kosterlitz-Thouless-like breaking of the 
electromagnetic $U_{em}(1)$ symmetry, and thus 
superconductivity, once a fermion 
mass gap opens up. The wavy line represents the $SU(2)$ 
gauge boson $B_\mu^3$,
which remains massless, while the blob denotes an insertion 
of the fermion-number
current  $J_\mu={\overline \Psi}\gamma_\mu \Psi$.
Continuous lines represent fermions.}}
\label{ktbreak}
\end{figure}
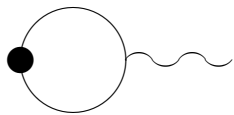
\end{centering}

\be
    {\cal S}^a = <B^a_\mu|J_\nu|0>,~a=1,2,3~; \qquad J_\mu ={\overline
\Psi}\gamma _\mu \Psi 
\label{matrix}
\ee
As a result 
of the colour group structure only the massless $B^3_\mu $ 
gauge boson of the $SU(2)$ group, corresponding to the $\sigma _3$
generator in two-component notation, contributes to the graph. 
The result is~\cite{RK,dor}:
\be
    {\cal S} = <B^3_\mu|J_\nu|0>=({\rm sgn}{M})\epsilon_{\mu\nu\rho}
\frac{p_\rho}{\sqrt{p_0}} 
\label{matrix2}
\ee
where $M$ is the parity-conserving fermion mass 
(or the holon condensate in the context of the 
doped antiferromagnet). 
In our case this mass is generated {\it dynamically} by means of the $U_S(1)$ interactions, as we discussed above, provided the coupling constants were lying in the appropriate
(strong) regime of the phase diagram of gig. \ref{pdiag}. 

The result (\ref{matrix2}) is {\it exact} in perturbation theory,
in the sense that the only modifications coming from higher loops 
would be a multpilicative factor $\frac{1}{1-\Pi (p)}$ 
on the right hand side, 
with $\Pi (p)$ the $B_\mu^3$-gauge-boson vacuum 
polarisation function~\cite{RK}.

As discussed in \cite{dor,RK}, 
the $B^3_\mu$ colour component
plays the r\^ole of the {\it Goldstone boson}
of the spontaneously broken fermion-number symmetry.
If this  symmetry is exact, then the gauge boson $B_\mu^3$ 
remains {\it massless}. 
This  
is crucial for the superconducting properties~\cite{dor}, given that 
this leads to the appearance of a {\it massless pole} in the 
electric-current two-point correlators, the relevant graph being depicted in 
figure \ref{pole}.

\begin{centering} 
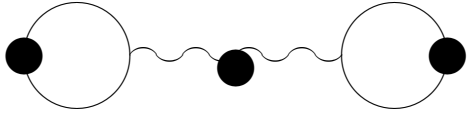
\begin{figure}[h]
\bigphotons
\begin{picture}(5000,5000)(0,0)
\put(15000,0){\circle{100000}}
\drawline\photon[\E\REG](17000,0)[8]
\put(13000,0){\circle*{1400}}
\put(\pmidx,-450){\circle*{1400}}
\global\advance\photonbackx by 2000
\put(\photonbackx,0){\circle{100000}}
\global\advance\photonbackx by 2000
\put(\photonbackx,0){\circle*{1400}}
\end{picture}
\vspace{3cm}
\caption{{\it The lowest-order
contribution to the electric current-current correlator $\langle
0|J_{\mu}(p)J_{\nu}(-p)| 0\rangle$. The
blob in the propagator for the gauge boson $B_\mu^3$ 
indicates fermion loop (resummed) corrections. 
The blob in 
each fermion loop indicates an insertion of the current $J_{\mu}$.}}
\label{pole} 
\end{figure}
\end{centering}

It can be shown~\cite{dor} that 
in the massive-fermion (broken $SU(2)$) 
phase, the effective low-energy theory 
obtained after integrating out the massive fermionic degrees of freedom
assumes the standard London action for superconductivity, the massless
excitation $\phi$ being defined to be the {\it dual} of $B_\mu^3$:
\be
\partial_\mu \phi \equiv \epsilon_{\mu\nu\rho}\partial_\nu B_\rho^3
\label{dual}
\ee
All the standard properties of superconductivity, Meissner effect (strongly type II~\cite{dor}), flux quantization and infinite conductivity, follow 
then in a standard way, provided the excitation $\phi$ (and, hence, $B_\mu^3$)
is exactly massless.

Having discussed the spontaneous breaking of fermion number symmetry 
in the massive fermion phase, it is natural to enquire on the 
nature of symmetry breaking, in the sense of establishing the 
existence or absence of a local order parameter. 
In this respect, our
discussion will parallel that of ref. \cite{dor}. 
The neutral parity-invariant condensate 
$<{\overline \Psi}_1 \Psi _1 - {\overline \Psi}_2 \Psi _2 > $,
generated by the strong $U_S(1)$ interaction, 
is {\it invariant} under the $U(1) \otimes U_E(1)$, as a result 
of the $\tau_3$ coupling of $B_\mu^3$ in the action, and hence 
does not constitute an order parameter. This is a characteristic
feature of our gauge interactions.
Putative charge $2e$ or $-2e$  order parameters, like the pairing interactions 
among opposite spins in the statistical model of \cite{fm,dor}, 
e.g. $<\Psi _1 \Psi _2>$, $<{\overline \Psi}_1 
{\overline \Psi}_2>$~\footnote{In four-cmponent notation, such
fermionic bilinears correspond to 
$<\Psi \gamma _5 \Psi >$, $<{\overline \Psi} \gamma _5 {\overline \Psi}>$,
considered in \cite{dor}.}
will vanish at any finite 
temperature, 
in the sense that strong phase fluctuations will destroy the 
vacuum expectation values of the respective operators, 
due to the Mermin-Wagner theorem. 
Even at zero temperatures, however, such vevs yield zero result 
to any order in perturbation theory trivially, due to the fact that 
in the context of the effective $B_\mu^3$ gauge theory of the broken 
$SU(2)$ phase, the gauge interactions preserve `flavour'. 
For a more detailed discussion  
on the symmetry breaking patterns 
of $(2+1)$-dimensional gauge theories,
and the proper definition of order parameter fields, 
we refer the reader to the 
literature~\cite{RK,dor}.

Thus, 
from the above analysis it becomes clear that 
gap formation, pairing and 
superconductivity can occur in the above model without implying any 
phase coherence. 

\section{Instantons and the fate of Superconductivity}

An important feature of our model 
is that, due to the  non-Abelian 
symmetry breaking pattern 
$SU(2) \rightarrow U(1)$, 
the abelian subgroup $U(1) \in SU(2)$,
generated by the 
$\sigma^3$ Pauli generator of $SU(2)$,
is {\it compact}, and 
may contain {\it instantons}~\cite{ahw}, which in three space-time dimensions 
are like monopoles, and are known to be responsible for giving 
a {\it small} but {\it non-zero mass} to the gauge boson $B_\mu^3$, 
\be 
        m_{B^3} \sim e^{-\frac{1}{2}S_0} 
\label{instmass} 
\ee
where $S_0$ is the one-instanton action, in a dilute gas approximation.
Its dependence on the coupling constant $g_2 \equiv g_{SU(2)}$ 
is well known~\cite{ahw}:
\be
     S_0 \sim \frac{{\rm const}}{g_2^2} 
\label{su2inst}
\ee
For weak coupling $g_2$ the induced gauge-boson 
mass can be very small. However, even 
such a small mass is sufficient to destroy superconductivity,  
since in that case there is no massless pole 
in the electric current-current correlator.

The presence of {\it massless} fermions, with 
zero modes around the instanton
configuration, 
is known~\cite{ahw} to suppress the instanton effects on the mass 
of the photon, and under certain circumstances, to be 
specified below, the Abelian-gauge boson may remain exactly massless
{\it even in the 
presence of non-perturbative effects}, thus leading to superconductivity,
in the context of our model. 
This may happen~\cite{ahw} 
if there are extra global symmetries
in the theory, whose currents 
connect the vacuum to the 
one -gauge-boson state, and thus they break spontaneously. 
This is precisely the case of the fermion number 
symmetry 
considered above. In such a case, 
the massless gauge boson is the Goldstone boson of the 
(non-perturbatively) spontaneously broken symmetry.

However, in our $SU(2) \otimes U_S(1)$ theory, 
discussed in this work,
as a result of the 
(infinitely strong) 
$U_S(1)$ interaction, a mass for the fermions is generated,
so we are not facing a problem with zero modes.
Our analysis is based on a Wilsonian treatment, where massive degrees of 
freedom are integrated out in the path integral. This includes the 
gapful fermions, and tha massive $SU(2)$ gauge bosons. 
The resulting theory, then, is a pure gauge theory 
$U(1) \in SU(2)$, and the instanton contributions to the mass of $B_\mu^3$
are present, given by (\ref{instmass}), in the one-instanton case.

We now remark that 
Supersymmetry is known~\cite{ahw} to suppress instanton contributions.
For instance, in certain $N=1$ supersymmetric models with massless fermions, 
considered in ref. \cite{ahw}  
the instanton-induced mass of the Abelian gauge boson is given by:
\be 
         m_{gauge~boson} \sim e^{-S_0} 
\label{instmass2}
\ee
which is suppressed, compared to the non-supersymmetric case (\ref{instmass}).

$N=2$ supersymmetric theories in three space-time 
dimensions
constitute additional examples of theories where the abelian gauge boson 
remains exactly massless, in the presence of instantons~\cite{ahw,dor2}. 
Such theories have complex representation for fermions, and hence 
are characterized by extra global symmetries (like fermion number). 
In view of our discussion above, such models will then lead to 
Kosterlitz-Thouless superconductivity 
upon gauging the fermion number symmetry. 

We also remark that 
in supersymmetric theories of the type considered 
here and in ref. \cite{diamand},
it is known~\cite{ahw} that  
supersymmetry cannot be broken, due to the fact that the Witten 
index $(-1)^F$, where $F$ is the fermion number, is always non zero. 
Thus, in supersymmetric theories the presence of instantons 
should give a small mass, if at all, in {\it both} the gauge boson and the 
associated gaugino, 
However, 
in three dimensional 
supersymmetric gauge theories 
it is possible that 
supesymmetry is broken by having the system in a `false' vacuum,
where the gauge boson remains massless, even in the presence of 
non perturbative configurations, while the gaugino acquires a 
small mass, through non perturbative effects. 
The life time, however, of this false vacuum is very long~\cite{ahw}, 
and hence superconductivity can occur, in the sense that 
the system will remain in that false vacuum 
for a very long period of time, longer than any other time scale 
in the problem.

The short reviews of symmetry-breaking patterns and the r\^ole of 
non-perturbative effects, just presented, 
provided us with the 
necessary equipment to attempt a construction of 
possible scenaria, which might  
simulate 
the interesting physics 
underlying the high-temperature superconducting cuprates.
A rather preliminary and heuristic discussion 
will be presented in the next section. A more detailed analysis,
especially in the context of the statistical models of ref. \cite{fm},
requires proper lattice 
simulations which automatically incorporate 
non perturbative 
effects. This, however, falls beyond the scope of the present paper.

\section{Application to the Physics of high-temperature superconductors} 

\subsection{Phenomenology of high-temperature superconducting materials}

\begin{figure}[htb]
\vspace{0.2in}
\begin{center}
\parbox[c]{4in}{\rotate {\rotate {\rotate{\psfig{figure=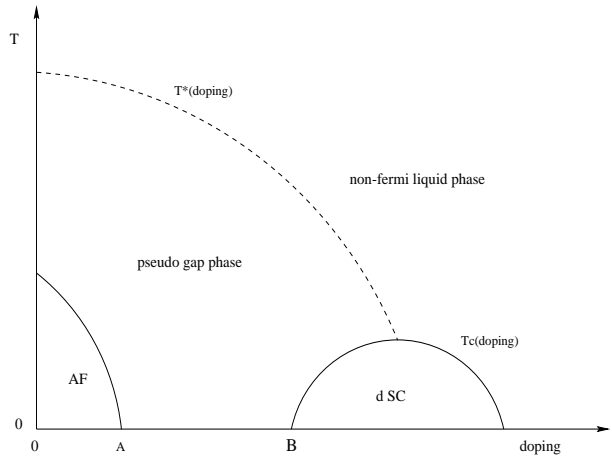,height=8cm,width=6cm}}}}}
\end{center}
\caption{{\it The temperature-doping phase diagram summarizes 
the current (experimentally observed) situation in high-temperature 
superconducting cuprates. Notice the existence of an intermediate 
zero-temperature phase, characterised by the existence of preformed pairs, leading to a pseudogap.}}
\label{pseudogap} 
\end{figure}

In this section we would like to consider a possible application of
the above $SU(2) \otimes U_S(1)$ model~\cite{fm} 
to the physics of high-temperature superconducting cuprates.
Recent experiments~\cite{underdoped} have demonstrated an 
extremely unconventional and rich structure of these materials, 
not in their superconducting phases, but rather in the normal phase. 
The phenomenology of the high-temperature cuprates may be summarized
by the 
temperature-doping concentration phase diagram, 
shown in figure \ref{pseudogap}.

The phase diagram shows clearly a very-low (including zero) doping 
antiferromagnetic phase (AF). Above a critical doping concetration (point $A$ in fig. \ref{pseudogap}), AF order is destroyed, but the interesting issue is the
existence of a phase, named `psuedo-gap phase', which interpolates 
between the AF and the sperconducting 
phases (dSC), the latter being known to be 
of $d$-wave type~\cite{tsuei}.  

It is a
general belief today, supported by 
many experimental results~\cite{underdoped},
e.g. results on optical conductivity, photo-emission, transport etc., 
that the pseudo-gap phase is characterised 
by pairing (`pre-formed pairs'), leading to the existence of a mass 
(pseudo)-gap 
in the fermionic spectrum, which however is not accompanied   
by phase coherence. This situation  
is in sharp contradction 
with 
in 
standard BCS theories of superconductivity, according to which 
phase coherence 
appears simultaneously with the appearance of a gap. 

The pseudogap phase is separated by a critical temperature curve
$T_c$(doping) from the $d$-wave superconducting state, characterized by 
the sharp drop in resistivity, and strong-type II superconductivity
(penetration depth of external magnetic fields is of order of a few thousands of Angstr\"oms). 
Today, the general belief is that the superconducting 
pairing is of BCS type involving four-fermion itneractions
among the charged excitations. However, the four-fermion 
interactions do not have to be phononic.

The pseudo-gap phase 
is also separated by another curve $T^*$(doping) from 
the normal state phase, where there is no gap,
but where there are abnormal normal state properties, such 
as linear dependence of the electrical resistivity with temperature 
for a wide range of temeprature scales
etc.
All such properties point towards a non-fermi liquid behaviour
of the normal state,
which is experimentally observed, as far as we understand, 
at least in the regime 
of optimum doping (shown in the figure \ref{pseudogap}).

\subsection{Strongly-coupled $U(2)$ gauge theory and 
the pseudogap phase}

In this section we shall argue that the gauge theory $SU(2) \otimes U_S(1)$
of ref. \cite{fm}, whose low-energy limit has been studied in this paper
in some detail, may provide a satisfactory qualitative 
explanation of the phase diagram of fig. \ref{pseudogap},
especially as far as the appearance of a pseudogap phase
is concerned. For the purposes of this article, 
we shall concentrate in the zero temperature 
region of the graph. Our method
will be that of ref. \cite{fmb}, 
i.e. approaching the pseudogap phase by studying the 
excitations about the nodes of the $d$-wave superconducting gap. 
We shall not deal here with excitations away from the nodes of the gap. 
Our hope will be that similar (long range) 
gauge-interaction phenomena are responsible 
for the formation of the bulk of the $d$-wave gap and the pseudogap. 
{}From the preliminary finite-temperature analysis of ref. \cite{fmb}
it becomes clear that the gaps that open up at the nodes 
disappear at much lower temperatures ($T < 0.1 K$) 
than the bulk of the $d$-wave type 
gap ($T_c = {\cal O}(100 K)$), and this means that the predictions
made in this work, if true, can be realised only if one looks at low
temperatures~\footnote{However, we point out that the presence of external 
magnetic fields may enhance these values~\cite{fmb}, as, for instance, 
is the case of 
the experiments
involving thermal conductivity measurements~\cite{ong}. See discussion 
at the end of this section.}. 

Before starting our 
analysis on the (zero-temperature) pseudogap phase,
we should point out that 
the gauge theory at hand, is also in agreement with deviation 
from fermi-liquid behaviour in the normal (no mass gap) phase. Indeed, 
as we discussed in previous articles~\cite{aitchmav},
a $U(1)$ fermion-gauge theory in $(2+1)$ dimensions, 
in which the mass of the fermions is generated
only dynamically through the gauge interactions,  
is characterised 
by non-trivial infrared fixed points, 
which according to general arguments~\cite{shankar} is sufficient to drive the theory away from the Landau fermi 
liquid (trivial infrared structure).

Let us continue our discussion 
on the diagram of fig. \ref{pseudogap}
by considering the 
Antiferromagnetic phase (AF).
In such a phase,  
the only excitations are assumed to be spin degrees of freedom. There 
are no charged excitations, and the pertinent dynamics is described by 
the magnon $z$ sector of the model (\ref{Hub}). 
As the doping exceeds a critical 
concentration (point A in fig. \ref{pseudogap}), the 
antiferromagnetic order is destroyed. 
In the context of simple low-energy $CP^1$ models, which describe adequately
the dynamics of the AF sector, this can be seen easily by applying 
renormalization-group arguments~\cite{rg,altaba}, and taking into account the 
dependence of the respective coupling constant on doping, in the 
way explained in refs. \cite{dor,dorstat}.

The important question is whether superconductivity does not set in immediately, but one has to pass through the intermediate phase $AB$, where 
a `pseudo-gap' appears, but no phase coherence exists.  
As we shall argue now, our strongly-coupled gauge theory $SU(2) \otimes U_S(1)$, presented above, may offer an explanation for the phenomenon.

To this end, we first remark that 
above the critical doping concentration that marks the on-set of 
disorder
($A$ in fig. \ref{pseudogap}),   
the $z$ magnons are massive, with masses which 
themselves depend on the doping 
concentration~\cite{altaba,dorstat,marchettilu}. 
In this regime, there are both charge and spin ($z$ field) excitations. 
Integrating out the massive magnons $z$,  
the long-wavelength dynamics 
of the charge excitations is described by the effective 
$SU(2) \otimes U_S(1)$
gauge theory of ref. \cite{fm}, (\ref{su2action}). 
The gauge $U_S(1)$ 
interaction is capable of 
inducing dynamical opening of a holon gap (pairing) 
if the pertinent coupling constant 
of the statistical model lies inside the $SU(2)$ broken 
regime of the phase diagram of fig. \ref{pdiag}. 

\begin{centering}
\begin{figure}[htb]
\epsfxsize=2in
\centerline{\epsffile{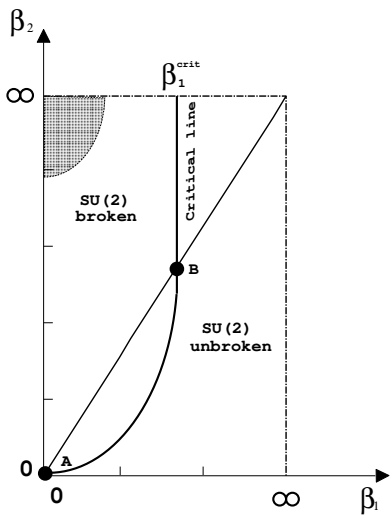}}
\vspace{1cm}
\caption{{\it Phase diagram 
for the $SU(2) \times U_S(1)$ gauge theory,
viewed as a low-energy continuum limit of 
the solid-state model 
for doped antiferromagnets of ref. [6]. 
The straight line indicates the specific relation 
of the coupling constants in the model.}}
\label{fig21fm}
\end{figure}
\end{centering}

We now remark that in the statistical model (\ref{Hub}), which 
will be the basis 
of our discussion in this section, the inverse couplings of the $SU(2)$ and $U_S(1)$ 
gauge groups lie in the straight line AB depicted in fig. \ref{fig21fm},
as a result of (\ref{connection}). In the condensed-matter model 
of ref. \cite{fm}, then, the local gauge group is $U(2)$ rather than 
$SU(2) \otimes U_S(1)$~\footnote{This, however, does not affect the 
results of the previous analysis pertaining to the mass generation.
The only difference of the $U(2)$ case is the {\it absence } of {\it baryons} 
from the spectrum~\cite{kawamoto}.}.

At present, the precise shape of the critical line 
is not known, since it requires the construction of an appropriate  
fermionic algorithm, which will allow for a proper lattice study of the model. 
The strong coupling analysis in this paper has  demonstrated, however, 
that the critical line passes through the origin of the graph,
concaving upwards in the way shown in the figure. 
Also, from the behaviour of the critical line about the 
point $(\beta_2=\infty, \beta_1^c)$,   
it is evident from the graph of figure \ref{fig21fm}
that the intersection point $B$ defines an
upper bound for the inverse coupling $\beta_1$,
in order for the system to be in the mass-generation phase. 
For all practical purposes it is qualtitative meaningful to 
assume an almost vertical shape of the critical 
line at the intersection point B, which implies that 
the critical coupling for mass generation for the coupling $\beta_1$ 
in the statistical model is still given by the single $U(1)$ gauge theory 
critical coupling, i.e. 
$\beta_1 < \beta_c^1$ (see figure \ref{pdiag}).
It is known that $\beta^c_1 \sim 32/\pi^2$~\cite{app,kocic}.
so, on account of our discussion in subsection 3.1 and 
(\ref{connection}), 
such a gauge pairing would occur in the following range of doping 
concnetrations:
\be 
      \delta _{AF} < \delta < \delta _c^{(2)} \equiv 1 - \frac{\pi^2}{32}\frac{\Lambda}{J} 
\label{pseudo}
\ee
where $\delta _{AF}$ denotes the doping concentration at which the AF order is destroyed.

This phase is characterised by the breaking of chiral symmetry.
However, as discussed in \cite{RK,dor}, 
and reviewed in section 4,  
the symmetry breaking 
occurs {\it without} a local order parameter. Strong phase fluctuations 
destroy the putative order parameter for $2+1$-dimensional $QED$.
This is an {\it exclusive feature} of the $2+1$ gauge interactions, and 
as we argue now, it is responsible for the appearance of a pseudo gap. 
{\it The concept of the pseudogap 
is associated precisely 
with the presence of a non-vanishing 
mas gap and thus the existence of `pre-formed pairs'
in underdoped cuprates~\cite{underdoped}, but in 
absence of  
a 
local order parameter (phase coherence) 
in the model}. The situation is analogous to 
the 
Kosterlitz-Thouless 
mode of symmetry breaking~\cite{KT}. 
The important issue to understand is why there is no superconductivity
in the model. 

This issue is related to the presence of instantons, discussed in the 
previous section,
which are responsible for giving a small but finite 
mass to the gauge boson $B_\mu^3$,
and thus destroying the basic criterion of superconductivity 
(see fig. \ref{pole}). 
The presence of fermions does not change this. 
In the broken $SU(2)$ (gapped) phase of the model (\ref{su2action}) 
the fermions are already massive, 
due to the extra $U_S(1)$ interactions, 
and hence there is no issue of zero modes
that could screen the instanton effects. 
{}From (\ref{instmass}), as well as the fact that the one-instanton action
exhibits the following dependance on the $g_{SU(2)}$ coupling constant
for the case at hand~\cite{ahw}:
\be
    S_0 \sim \frac{{\rm const.}}{g_{SU(2)}^2} \qquad ; \qquad g^2_{SU(2)} \propto J(1-\delta) 
\label{instcoupl}
\ee
one observes that the instanton-induced $B_\mu^3$-boson mass 
decreases upon increasing the doping concentration $\delta$ 
in the sample.

In the context of the statistical models of ref. \cite{fm}, etc, 
one should also consider the coupling of superconducting planes,
by means of Coulomb interactions among the charge carriers (electrons). 
Such interactions may result in a small leakage of electrons 
across the planes, which inevitably leads to fermion-number 
non-conservation on the plane. In ref. \cite{diamand}, 
within a spin-charge separating framework, 
such an interplanar coupling has been represented by inserting in the 
path-integral a 
term of the form:
\be
    \int d\eta e^{2i \int d^3x {\overline \eta} \Psi^\alpha z_\alpha + h.c. }
\label{coupling} 
\ee
where $\alpha=1,2$ runs over `colours' in the model of $SU(2)$, 
and 
$\eta$ is a Majorana spinor. Due to this, the term (\ref{coupling}) 
in the effective action violates fermion number, and is interpreted 
as implying a hopping of {\it both} spin ($z$) and charge ($\Psi$) 
degrees of freedom. Notably, $\eta$ may 
play the r\^ole of the supersymmetric partner 
of the $B_\mu^3$ gauge boson, in a $N=1$ supersymmetric 
formulation of the model~\footnote{This 
supersymmetry
carries non-trivial dynamical information about 
the spin-charge separation mechanism underlying the model,
and hence it is different from the non-dynamical global supersymmetry
algebras, 
at specific points 
of the coupling constants, discovered in \cite{sarkar}.}, 
which is possible upon certain relation~\cite{diamand} 
among the couplings of the microscopic spin-charge separating model 
of \cite{fm}.

The explicit breaking of the fermion number symmetry 
by the interplanar coupling, as well as the absence of fermion 
zero modes in the massive phase
(due to the $U_S(1)$ interactions) imply that the presence of fermions 
will not cancel the instanton-induced small mass of the gauge boson 
(\ref{instmass})~\footnote{Although, a reduction of  order (\ref{instmass2}) 
might be expected in $N=1$ supersymmetric 
cases~\cite{diamand}, occuring for particular values of doping.
See discussion below.}.  
In such a case, then, the gapped phase will be characterised by 
the presence of pairing, mass gap, but no phase coherence and 
superconductivity, features shared by the pseudogap phase 
observed in cuprates (fig. \ref{pseudogap}). 

The above considerations are rather heuristic
at present. The complete analysis 
would necessitate a lattice simulation of the model (\ref{action})
in the presence of instanton effects (in the broken phase after mass generation
due to $U_S(1)$ interactions). 
Analytical results at present exist only for $N=2$ supersymmetric theories,
as we mentioned above~\cite{ahw,dor2}, which however seem not to
correspond to the physics of the 
cuprates~\cite{diamand}, which appear to have at most 
$N=1$ supersymmetry upon coupling of the superconducting planes. 
Supersymmetry supresses instanton effects in some cases (\ref{instmass2}),
and some times may lead to a massless gauge boson, although such a case 
at present seems to characterise $N=2$ supersymmetric theories. 
We hope to return to a more detailed study 
of such issues 
in the future.

The suppression of instanton effects by supersymmetry, which in our 
class of statistical models may occur for certain doping 
concentrations~\cite{diamand} points to the following possibility: 
As one increases the doping concentration,
a region is reached where there is a special relation among the various 
coupling constant of the effective spin-charge separating theory, 
leading to a $N=1$ supersymmetry~\cite{diamand}
For instance, in the context of models of ref. \cite{dorstat} such a 
supersymmetric point could be reached 
for doping concentrations $\delta ^*$, such that 
$t' \sim \sqrt{JJ'} (1-\delta)^{3/2}$, where the prime denotes
next-to-nerarest neighbor hopping ($t$) and Heiseneberg exchange
energies ($J$). 
By tuning the couplings one may arrange- always in the cotnext 
of phenomenological models- for a situation 
in which $\delta ^* \le \delta _c^{(2)}$.
This would imply that, within the region of dopings for which 
the gauge statistical 
interactions are responsible for the opening of a gap and 
pairing,  
the suppression of instanton effects due to the presence 
of (supersymmetric) fermions may be sufficient to 
allow for a gauge-theory-induced 
Kosterlitz-Thouless (KT) superconducting gap at the $d$-wave nodes. 
The KT nature of the gap implies that once opened such a gap cannot affect 
the $d$-wave character. This scenario for superconductivity 
has been advocated in ref. \cite{fm}. 
In a related, but less probable,  
scenario,
the tunnelling to a 
`false' supersymmetry-broken 
vacuum~\cite{ahw} could 
occur in the $dSC$ region of the phase diagram (see fig. \ref{pseudogap}),

\subsection{Superconducting phase and additional four-fermion interactions}

Despite these appealing 
scenaria for the r\^ole of gauge (spin-spin) interactions 
for inducing KT superconducting gaps at the nodes, 
in the realistic situation 
the onset of ($d$-wave) superconductivity occurs 
at higher doping 
concentrations, for which 
the attractive 
four-fermi couplings among charged excitations 
in the effective fiedl theory 
become strong enough, so as to overcome the gauge interactions,
and lead to a standard BCS type pairing among the charged excitations.

In such a case  
one then is forced to 
consider the effect of additional contact interactions, among the {\it holons}, which are up and above 
the gauge interactions considered so far. The simplest, and most likely
the most relevant, of such interactions are contact {\it four fermion}
interactions~\cite{semenoff,dor}. From a microscopic point of view such interactions 
may arise 
as an effective way to describe the tendency of holes to 
occupy nearest neighbor lattice sites~\cite{dor,Sha},
or in some recent scenaria 
they may describe attractions due to 
screened Coulombic interactions
among charged excitations, as a 
result of the interplanar coupling~\cite{leggett}.  
The corresponding coupling constant will again depend on the doping concentration~\cite{dor}, and this will imply interesting phase structures.

The additional four-fermion 
interactions are, then, 
viewed as being responsible for the appearance of an 
ordered $d$-wave state, with the gap being characterized 
by four nodes. 
We shall be fairly phenomenological, 
given that a detailed incorporation of extra four-fermion 
interactions in our strongly-coupled
gauge model is not striaghtforward. However, a phenomenological analysis will 
be sufficient to demonstrate the main unconventional features of our scenario for 
high-temperature 
superconductivity, at least in the context of a continuum  
effective field theory.  

It is known that, in the context of relativistic models we consider here,
as a result of linearization about the nodes of a 
$d$-wave superconducting gap, the four-fermion interactions are
the only ones which become renormalizable (relevant) 
in the $1/N$ framework, where $N$ 
is a flavor number for fermions. 
As an instructive example, 
consider, for instance, Gross-Neveu type four-fermion couplings
in the effective lagrangian~\cite{fkm2}: 
\be
L_{4f} = \kappa \sum_{a=1}^{2} \left( {\overline \Psi}_a \Psi ^a  \right)^2  
\label{ffermi}
\ee
where $\Psi _a$ are the relativistic spinors (\ref{twospinors}) 
describing the excitations about the nodes of a $d$-wave gap.

{}From standard arguments~\cite{gat} on the phase structure of Gross-neveu 
type couplings, we are considering here, 
we know that pair formation, and hence mass generation, 
in four-fermion theories occurs for dimensionless inverse couplings, 
$\lambda \equiv \frac{1}{2\kappa \Lambda} $
weaker than a critical value,
$2/\pi^2$. 
However, the full phase diagram, incorporating 
the $SU(2)$ and $U_S(1)$ couplings as well, 
as appopriate for the model of \cite{fm}, 
will be more complicated. 
However, for our purposes in the present work 
it will be sufficient to 
consider only the effects of the $U_S(1)$ coupling, 
responsible for the mass generation in the model. 

\begin{figure}[htb]
\vspace{0.2in}
\begin{center}
\parbox[c]{5in}
{\psfig{figure=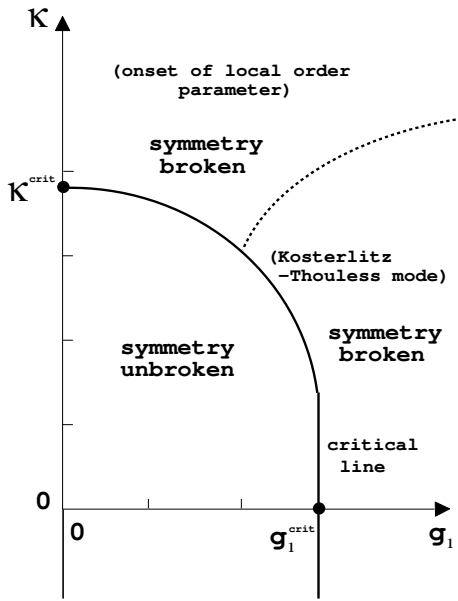,height=8cm,width=6cm}}
\end{center}
\vspace{1cm}
\caption{{\it A generic phase diagram of the theory with $U_S(1)$ gauge 
and four-fermion 
(Gross-Neveu) interactions. The critical line separates the phase of unbroken symmetry from that of broken symmetry. The symmetry breaking is due to 
the fermion condensate. The dotted line is conjectural at present,
and indicates the on-set of a local order parameter due to the dominance 
of four-fermion interactions.}}
\label{fourferd}
\end{figure}

A phase diagram for an
abelian gauge theory with extra four-fermion interactions (of Gorss-Neveu type)
has been  
derived analytically 
in the context of a Schwinger-Dyson large fermion-flavour analysis
in ref.~\cite{dor},
and in principle the result can be checked in lattice models.
For our purposes in this work it will be sufficient to 
assume the validity of the large-flavour-number 
continuous results of ref. \cite{dor}, and concentrate on the pertinent 
phase diagram
in the coupling constant 
space between the gauge $g$ and four-fermion couplings $\kappa$,
depicted in fig. \ref{fourferd}.

In toy models of doped antiferromagnets~\cite{Sha,dorstat}, 
which are sufficient for our
illusttrative purposes, 
such Gross-neveu four fermion terms are expressing 
the tendency of holes to break as less bonds as possible 
in the antiferromagnetic lattice, which is the configuration 
featuring the holes sitting next to each other. 
Such interactions, may be described by adding to the Hamiltonian 
{\it attractive} 
four-fermion interactions of the form:
\be
-\kappa :\psi_1^\dagger \psi_1(j)::\psi_2^\dagger \psi_2(j+1):  
\label{shaff}
\ee
where $\psi_\alpha, \alpha=1,2$ are Grassmann (holon) 
operators, $j$ denote lattice site, and $: \dots :$ denotes normal ordering 
of quantum operators. The normal ordering conventions
are such that a fermion bilinear is written as:
\be
     \psi^\dagger \psi = :\psi^\dagger \psi : + < \psi^\dagger \psi>, 
\qquad <\psi^\dagger \psi >=\delta 
\label{no}
\ee
with $\delta$ the doping concentration in the sample.   
Such terms may be asssembled, in the continuum, low-energy, 
limit~\cite{dorstat,dor},
into Gross-Neveu four-fermion terms of the form (\ref{ffermi}), where the 
spinors are constructed as in (\ref{twospinors}). 

At this stage, the coupling constant $\kappa$  
is a phenomenological parameter. However, 
from quite generic arguments, 
one would expect  
it to 
increase upon increasing the doping concentration $\delta$ in the sample,
since the larger the doping, the bigger the probability 
of the holes to lie in adjacent sites of the lattice. 

At present, the only case where 
four-holon-operators appear with well-defined coupling constants 
in terms of 
the microscopic parameters of the theory,
is the $t-j$ or Hubbard model case, 
where, however, the four fermion interactions are 
{\it repulsive}~\cite{dorstat,semenoff}. 
In the models of ref. \cite{dorstat}, for instance, 
such Hubbard four-fermion couplings   
$     \kappa_{Hubbard} $ assume the generic form: 
\be 
     \kappa_{Hubbard} \propto 	\kappa _0 (t', J') \frac{1}{1-\delta} 
\label{kappa0}
\ee
where $\kappa _0 (t',J')$ is an appropriate function of the 
next-to-nearest hopping element, and Heisenberg exchange energies. 

Combining (\ref{shaff}) with such repulsive interactions, one 
then arrives at a generic coupling for four-holon (Gross-Neveu) operators 
in the model of the form: 
\be
     \kappa_{4f} = \kappa (\delta ) - \kappa _0 (t', J') \frac{1}{1-\delta} 
\label{finalkappa}
\ee
where $\kappa (\delta)$ is, at present, a phenomenological parameter, 
which however is expected to increase, as we said, 
with increasing $\delta$. 
For the four-fermion interactions  to be 
attractive one needs $\kappa_{4f} >0$, and this places 
restrictions on the regime of doping, for which the 
interactions are going to lead to dynamical mass gap generation.

When combined with the phase diagram of fig. \ref{fourferd}, 
this implies that pairing due to four-fermion interactions 
would occur for 
doping concentrations in 
a 
region determined by the critical line of fig. \ref{fourferd}. 

By appropriately choosing $\kappa _{4f}$, in
the context of phenomenological models, 
it is then possible to arrange for a situation like 
the one depicted in 
fig. \ref{pseudogap}, where 
the zero-temperature pseudogap phase interpolates between the AF 
and the standard BCS type $d$-wave superconducting theory. 

Notice that the dynamical mass generation due to four-fermi couplings
leads to a {\it second order} transition, at zero temperatures, 
and hence to phase coherence,
as is standard in BCS type of pairing. This should be contrasted 
with the gauge situation described above, which leads to Kosterlitz-Thouless
type of breaking~\cite{RK,dor,fm}. 
It would probably imply the existence of a cross-over 
line (dotted) in the diagram of figure 
\ref{fourferd}, separating the region of the broken symmetry phase where a local order parameter is present, due to the dominance of the four-fermion itneractions, from the region where the Kosterlitz-Thouless mode 
of symmetry breaking (absence of a local 
order parameter) occurs, due to the dominance of the 
gauge interactions. Such phase diagrams should be confirmed 
by detailed lattice simulations, using appropriate fermionic algorithms, 
which fall beyond the 
scope of the present work.

The interesting feature is that {\it experimentally} 
one can make a distinction 
between a gap induced by the gauge interactions, or by four fermi 
interactions, as a result of different scaling of the mass gap with 
an externally applied magnetic field. 
A suggestive experiment along these lines is 
that 
of ref. \cite{ong}, measuring the behaviour of the thermal conductivity,
in both the superconducting and  `pseudo-gap' phases.  
Details on such issues are discussed in refs.~\cite{fmb,fkm,shov,fkm2}.

\section{Conclusions} 

In this work we have described a strong
coupling expansion
for an $SU(2) \otimes U_S(1)$ gauge theory, in three dimensional
space time.
{}From the physical point of view, such models 
may serve either as a prototype for physical applications 
to the physics of excitations about nodes in a $d$-wave 
high-temperature superconductor~\cite{fm}, or - when formulated 
in Euclidean lattices - as describing high-temperature phases 
of four-dimensional gauge theories of the Early Universe. 

Our analysis has indicated a phase diagram of the form 
depicted in figure \ref{pdiag}. 
The analytical results obtained in the present 
article pertained to 
the strongly coupled $SU(2)$ sector. We have shown the absence of a
finite critical coupling for $SU(2)$ symmetry breaking. 
The breaking of $SU(2)$ 
is induced by dynamical parity-conserving 
mass generation due to the $U_S(1)$ strongly-coupled sector. 
The shape of the critical line, spearating the phases
of broken $SU(2)$ symmetry, is still conjectural, since it requires 
proper lattice simulations with dynamical fermions, which is under
consideration at present. 

An important ingredient in our analysis, which was motivated from 
our condesed-matter ancestor models~\cite{fm}, was the use of 
``naive'' Dirac spinors on the lattice, and {\it not} Wilson fermions. 
The latter are known
to violate explicitly parity-symmetry breaking, due to the Wilson term.
This may lead to spontaneous violation of parity symmetry~\cite{aoki,parlat}, 
and therefore to a completely different phase diagram, although the issue is 
still unsettled~\cite{bitar}~\footnote{A similar effect may be induced by 
external electromagnetic interactions, which from a 
condensed-matter point of view are natural to consider.
Recently, the effects of constant magnetic fields on 
the opening of a mass gap at the nodes of d-wave 
high-temperature superconductors have been considered in 
the experiments of~\cite{ong}. 
Claims that this may induce, for strong enough fields, 
a change of state of the condesate into a parity-violating one,
have been made~\cite{laughlin}. 
Indeed, 
in the case 
of a constant external 
magnetic field, perpendicular to 
the spatial plane, one 
has an external source term violating parity and time-reversal symmetry. 
For strong enough source fields 
it is possible that a parity-violating
condensate
is magnetically induced. Such a phenomenon 
is at present a conjecture, which 
needs to be demonstrated 
analytically (via Schwinger-Dyson analysis) 
or on the lattice. We postpone such an issue for future investigations. 
We should mention, however, 
that recent preliminary lattice~\cite{fkm} or continuum~\cite{fmb}
analyses, in the presence of an external field, 
showed that  the magnetic field
enhances the 
parity-conserving condensate.}. 

An interesting application of our strongly-coupled gauge 
theory $SU(2) \otimes U_S(1)$ was argued to be provided by a 
possible explanation of the (zero-temperature)
pseudo-gap phase between the antiferromagnetic and $d$-wave 
superconducting phases of the high-temperature cuprates 
(see fig. \ref{pseudogap}). 
Due to the special symmetry breaking patterns, $SU(2) \rightarrow U_(1)$ 
in the phase where a fermion mass is generated dynamically 
by the $U_S(1)$ statistical interactions in the model~\cite{fm}, 
and the existence of instanton configurations in the compact 
$U(1) \in SU(2)$, a small mass for the $U(1)$ gauge boson can be generated.
Such a small mass, although does not prevent pairing, however it spoils superconductivity, since it leads to the disappearance of the massless pole in the 
electric current two-point correlators. 
As explained in the text, in such a sector, the gauge theory is responsible only for the opening of a mass gap in the fermion spectrum. However,
the mode of symmetry breaking is of Kosterlitz-Thouless type, and hence 
no order parameter exists, since strong phase fluctuations destroy it
(the parity-invariant fermion mass gap in the model of refs. \cite{fm}
is not an 
order parameter, since it is invariant under the respective symmetries,
and hence there is no contradiction with the Mermin-Wagner theorem). 
Such a KT mass gap
may be viewed as a pseudo-gap. In our scenario above, 
such a gap owes its presence 
to 
relativistic fermions at specific points of the fermi surface 
(e.g. nodes of the $d$-wave gap), argued to play a crucial  
r\^ole 
in the 
pertinent phase. 
We stress once more, that the important 
point in our approach~\cite{fm} was the 
Kosterlitz-Thouless nature of the gauge symmetry 
breaking induced by gauge interactions in $(2+1)$ dimensions~\cite{RK,dor},
which discriminates our approach from others~\cite{marchettilu,fisher}. 

Our belief that spin-gauge interactions may a play a 
crucial r\^ole in the underdoped and normal phase of the high-$T_c$ cuprates 
is strengthened by the abnormal properties of these phases~\cite{underdoped},
including the explicit observation of phase separation 
in the so-called stripe phase, occuring for a particular doping 
concentration~\cite{stripes}.
In the $d$-wave superconducting phase, four-fermion BCS-like pairing 
may indeed occur, although the attractive four-fermion interactions, 
most probably, are not due to phonons, but of electronic 
origin~\cite{leggett}. 

In the presence of {\it both} types of interactions, gauge and four-fermion,
the effective theory model presents interesting phases. 
One way to determine the origin of the dynamically-induced mass gap 
in the various pahses 
is to study the behaviour of the system under the influcence 
of external fields, like in the experiment of \cite{ong}. 
It is known~\cite{fkm,fmb,fkm2} that the gauge-field induced mass gap 
scales differently with an applied magnetic field as compared to the gap induced by 
four-fermi Gross-Neveu type interactions.  
Such a scaling may be determined 
by studying the 
thermal conductivity in the presence of 
an externally 
applied magnetic field, as in the experiments of \cite{ong}. Details of this 
analysis appear in~\cite{fkm2}. 

We are, of course, 
aware that the simple, effective gauge field theory analysis 
we have just presented, may not be sufficient to explain 
quantitatively the rich phenomenology of the high-temperature 
cuprates. We believe 
however, that it constitutes a 
step in this direction. Our hope is that,  
due to the 
simplicity and universality that underlies the  superconducting models
based on the gauge symemtry approach,
our 
results
capture essential features of the physical mechanism(s) underlying 
various phases of high-temperature superconductors, 
in much the same way as the single phonon BCS theory 
describes adequately the complicated physics of 
phononic superconductors. 
Moreover, as particle theorists, 
we also find this exercise  
very interesting, since it may imply that certain phenomena,
characterizing the Physics of the Early Universe, may have interesting 
counterparts in condensed-matter physics, and in particular the 
high-temperature (antiferromagnetic) superconductors. 
In this respect, 
specific mention should be made again to the work of Volovik~\cite{volovik}, 
who pursues the analogies between Particle and Condensed matter Physics, 
by suggesting solid-state experiments, involving superfluid Helium, 
as possible laboratory experiments which might shed light to the 
Physics of the early stages of our Universe. In the same spirit, the 
rich, and unconventional, 
structure of the high-temperature cuprates, depicted in fig. 
\ref{pseudogap}, may also find interesting, and possibly new, 
applications to Particle Physics. 
 
\section*{Acknowledgements}

It is a pleasure to acknowledge discussions with I.J.R. Aitchison,
J. Betouras, C.P. Korthals-Altes, M. Teper and A. Tsvelik.  
The work of D. Mc Neill is supported by a P.P.A.R.C. (UK) studentship.
K.F. wishes to thank the TMR project FMRX-CT97-0122 for partial 
financial support, and the Department of Theoretical Physics 
of Oxford University for the hospitality during the last stages of this work.

\newpage

\section*{Appendix 1: Rules for Multiple `Meson' Pairing}

When we come to calculate the meson pairs on a particular site there
is a subtle complication if there are two pairs on the site, ie. four
$\Psi$s and four $\overline{\Psi}$s. Let us take as an example the
function $[tr(\overline{A}_{i,\mu}A_{i,\mu})]^4$ which lies on the
link $(i, i+\mu)$. Expanded out this gives

\bea
&[\overline{\Psi}_{i}^{b,\alpha}
(-\gamma_\mu)^{\alpha\beta}\Psi_{i+\mu}^{a,\beta}
\overline{\Psi}_{i+\mu}^{a,\gamma}
(\gamma_\mu)^{\gamma\delta} \Psi_{i}^{b,\delta}]
\times
[\overline{\Psi}_{i}^{d,\epsilon}
(-\gamma_\mu)^{\epsilon\zeta}\Psi_{i+\mu}^{c,\zeta}
\overline{\Psi}_{i+\mu}^{c,\theta}
(\gamma_\mu)^{\theta\eta} \Psi_{i}^{d,\eta}]
\times \nn \\
&[\overline{\Psi}_{i}^{f,\kappa}
(-\gamma_\mu)^{\kappa\lambda}\Psi_{i+\mu}^{e,\lambda}
\overline{\Psi}_{i+\mu}^{e,\pi}
(\gamma_\mu)^{\pi\rho} \Psi_{i}^{f,\rho}]
\times
[\overline{\Psi}_{i}^{h,\sigma}
(-\gamma_\mu)^{\sigma\tau}\Psi_{i+\mu}^{g,\tau}
\overline{\Psi}_{i+\mu}^{g,\phi}
(\gamma_\mu)^{\phi\chi} \Psi_{i}^{h,\chi}]
\eea
Before, when we just had a term which produced one pair of mesons on each
site there was only one 
unique way of combining the $(\Psi\overline{\Psi})$ at that site to
make the meson. However now, as we see,
there are three ways at each site. Writing down the fields at site $i$;
\be
\Psi_{i}^{b,\delta} \overline{\Psi}_{i}^{b,\alpha} 
\Psi_{i}^{d,\eta}\overline{\Psi}_{i}^{d,\epsilon}
\Psi_{i}^{f,\rho}\overline{\Psi}_{i}^{f,\kappa}
\Psi_{i}^{h,\chi}\overline{\Psi}_{i}^{h,\sigma}
\ee
these can be combined into pairs in three ways
\bea
\Psi_{i}^{b,\delta} \overline{\Psi}_{i}^{d,\epsilon}
\Psi_{i}^{d,\eta} \overline{\Psi}_{i}^{b,\alpha}
&\times&
\Psi_{i}^{f,\rho} \overline{\Psi}_{i}^{h,\sigma}
\Psi_{i}^{h,\chi} \overline{\Psi}_{i}^{f,\kappa}  
\nn \\
or\;\;\;\Psi_{i}^{b,\delta} \overline{\Psi}_{i}^{h,\sigma}
\Psi_{i}^{h,\chi} \overline{\Psi}_{i}^{b,\alpha} 
&\times&
\Psi_{i}^{f,\rho} \overline{\Psi}_{i}^{d,\epsilon}
\Psi_{i}^{d,\eta} \overline{\Psi}_{i}^{f,\kappa}
\nn \\
or\;\;\;\Psi_{i}^{b,\delta} \overline{\Psi}_{i}^{f,\kappa}
\Psi_{i}^{f,\rho} \overline{\Psi}_{i}^{b,\alpha} 
&\times&
\Psi_{i}^{h,\chi} \overline{\Psi}_{i}^{d,\epsilon}
\Psi_{i}^{d,\eta} \overline{\Psi}_{i}^{h,\sigma}
\eea
when faced with a
choice of three possible meson states, we must
assume that each is likely to happen with equal probability (in a
sense they can be seen as quantum states with equal energy) and so we
write the final combination state at site $i$ as

\bea
&\frac{1}{3} 
\Psi_{i}^{b,\delta} \overline{\Psi}_{i}^{d,\epsilon}
\Psi_{i}^{d,\eta} \overline{\Psi}_{i}^{b,\alpha} 
\Psi_{i}^{f,\rho} \overline{\Psi}_{i}^{h,\sigma}
\Psi_{i}^{h,\chi} \overline{\Psi}_{i}^{f,\kappa}  +
\frac{1}{3} 
\Psi_{i}^{b,\delta} \overline{\Psi}_{i}^{h,\sigma}
\Psi_{i}^{h,\chi} \overline{\Psi}_{i}^{b,\alpha} 
\Psi_{i}^{f,\rho} \overline{\Psi}_{i}^{d,\epsilon}
\Psi_{i}^{d,\eta} \overline{\Psi}_{i}^{f,\kappa}
& \nn \\
&+ \frac{1}{3}
\Psi_{i}^{b,\delta} \overline{\Psi}_{i}^{f,\kappa}
\Psi_{i}^{f,\rho} \overline{\Psi}_{i}^{b,\alpha} 
\Psi_{i}^{h,\chi} \overline{\Psi}_{i}^{d,\epsilon}
\Psi_{i}^{d,\eta} \overline{\Psi}_{i}^{h,\sigma},
&
\eea
or
\be
\left\{
\frac{1}{3} tr_c[M_i^{\delta\epsilon}M_i^{\eta\alpha}]
tr_c[M_i^{\rho\sigma}M_i^{\chi\kappa}] + 
\frac{1}{3} tr_c[M_i^{\delta\sigma}M_i^{\chi\alpha}]
tr_c[M_i^{\rho\epsilon}M_i^{\eta\kappa}] +
\frac{1}{3} tr_c[M_i^{\delta\kappa}M_i^{\rho\alpha}]
tr_c[M_i^{\chi\epsilon}M_i^{\eta\sigma}] \right\}  
\ee
This must be contracted with the gamma matrices on the link and then
with the meson states on site $i+\mu$, calculated in the same way.

This procedure must also be followed when the four
$(\Psi\overline{\Psi})$ pairs on the site come from two functions on adjacent
sides meeting at that site.

\newpage

\section*{Appendix 2: The `Meson' Jacobian}

\pr
In this appendix we shall calculate the Jacobian 
of the transformation from the fermion variables 
${\overline \Psi}_{\alpha,a}, \Psi_{\alpha,a}$,
to the meson variables ${\cal M}_{\alpha\beta}^{ab} \equiv {\Psi}_{\beta}^b 
{\overline \Psi}_\alpha^a. $
This change of variables 
implies the following transformation in the path integral:
\be
\prod_i \int d \overline{\Psi}_i d \Psi_i \mapsto \prod_i \int d M_i,
\ee
we adapt the method outlined in \cite{kawamoto}. Let our initial path
integral be written;

\be
\prod_i \int d \overline{\Psi}_i d \Psi_i e^{S_{eff}[M_i]} = \prod_i
e^{S_{eff}[\frac{\delta}{\delta J}]} \int d \overline{\Psi}_i d \Psi_i
e^{tr JM_i} |_{J=0},
\label{sou}
\ee
\noindent
the above fermionic integral can be evaluated;

\be
\int d \overline{\Psi} d \Psi e^{tr JM} = \int d
\overline{\Psi} d \Psi e^{\overline{\Psi} J \Psi} = det J \;\;\;;\;\;\;
\overline{\Psi} J \Psi = \overline{\Psi}^{a \alpha} J^{ab\alpha\beta}
\Psi^{b \beta},
\ee
\noindent
where $det J$ means the product of the eigen-values of $J$, which is a
$2 \times 2$ matrix in two spaces.
However
there is a problem defining exactly what we mean by the measure $dM$,
since $M$ lives in two spaces. However if we instead regard $M$ as a
$4 \times 4$ matrix we can use some simple results from group
integration \cite{creutz,kawamoto} to define a measure. The following
identity is true for a $U(4)$ group integral
\be
\int dU \frac{1}{det RU} e^{tr J^{'}RU} = det J^{'},
\label{jprime}
\ee
where $R$ is any positive-definite hermitian matrix. The matrix
$J^{'}$ is chosen to be the equivalent matrix in $4 \times 4$ space to
the $J$ defined in (\ref{sou}) which lived in two $2 \times 2$
spaces. But the determinant of both are the same, $det
J^{'} = det J$. It is then a simple matter of associating, in
(\ref{jprime}), $dU \mapsto dM$ and $RU \mapsto M$ and we get
\be
\int d \overline{\Psi} d \Psi e^{tr JM} = \int dM \frac{1}{detM}
e^{tr JM},
\ee
\noindent
and our path integral becomes

\be
\prod_i \int dM_i e^{-\sum_i ln det M + S_{eff}}\;\;.
\ee

Our path integral, $dM_i$, over the $4 \times 4$ space which we
have represented as a group integral, can be
viewed as a multi-component generalisation of a contour
integral.

If $M_i$ had been a complex number, which is the case of a $U(1)$ 
gauge theory with Kogut-Susskind lattice fermions, we
would have had~\cite{kawamoto} 
\be
\int d\overline{\Psi}\Psi e^{\overline{\Psi}J\Psi} = J = \oint
\frac{dz}{2\pi i z}\frac{e^{Jz}}{z},
\label{cauchy}
\ee
This contour integral can be evaluated by parameterizing
$z=Re^{i\theta}$, which is of course a representation of the $U(1)$ group.
In such a case, 
this property of the Jacobian, expressible 
in terms of contour Cauchy integrals, 
can be used to infer  
stability of 
the broken vacuum
in strongly-coupled gauge theories, despite the appearance of a 
local maximum in the potential.
A detailed discussion 
may be found in \cite{kawamoto},
where we refer the reader for more details. 
Below we shall only concentrate to 
a description of the basic results, pertinent for our analysis  
in this work.

\begin{figure}[htb]
\vspace{0.2in}
\begin{center}
\parbox[c]{3in}{\psfig{figure=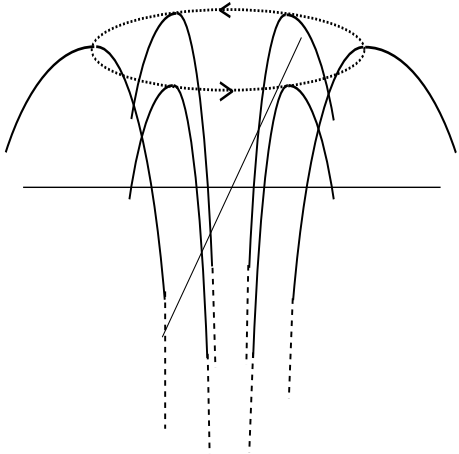,height=6cm,width=6cm}}
\end{center}
\caption{{\it Explanation on stability despite the existence of a local maximum
in the effective potential.}}
\label{effpot2f} 
\end{figure} 

In our problem, $M$ is a $4\times 4$ matrix and thus has
16 degrees of freedom. These degrees of freedom are illustrated in the
expansion of $M$ (\ref{mesontripl}) in terms of the ``lengths'' (the
${\cal A}$s and ${\cal F}$s) along
the 16 ``axes''. However we were interested in symmetry breaking along
just one of these ``axes'', $\sigma_3.{\bf 1}$, with a ``length'' given by
the complex number ${\cal A}_3$.
So we can view the important part of the integral $dM$ as being in the
complex plane
where the only degree of freedom is ${\cal A}_3$. 
In this respect 
we have complexified the V.E.V. of ${\cal A}_3$ 
in order to apply the contour integration. Eventually, 
the effective
potential will be minimized for  real $<{\cal A}_3>$.  
We have thus a
standard contour integral as above (\ref{cauchy}).

To make things clearer it is useful~\cite{kawamoto} to 
add an explicit chiral-symmetry-breaking fermion mass term, $\sum_i
m\overline{\Psi}_i \sigma_3 
\Psi_i$, to our lagrangian. 
The VEV
of ${\cal A}_3(i)$ can be written as~\cite{kawamoto}:
\bea
<{\cal A}_3(i)> &= Ue^{i\varphi}, &even \;sites, \nn \\
<{\cal A}_3(i)> &= Ue^{-i\varphi}, &odd \;sites,
\eea
The VEV of the mass term
$<m\overline{\Psi}_i \sigma_3  
\Psi_i>$ is $m<{\cal A}_3(i)>$ and so summed over even and odd sites
will pick out the real part of the complex number:

\be
\sum_i m<{\cal A}_3(i)> =\sum_i \frac{1}{2} mU\cos \varphi\;\;.
\ee
Our effective potential now with the mass term is given in the form

\be
V_{eff} \sim \frac{2}{3} ln U^4 - P(U^4) - \frac{mU}{6} \cos \varphi ,
\ee
where $P(U^4)$ represents the polynomial in $U^4$ which makes up the
rest of $V_{eff}$, given in the text (\ref{zeroeff}),(\ref{twoeff}). 
It is now clear how the apparent maximum at the
stationary point is interpreted. 
Since the path integral is
effectively a contour integral, we can 
choose 
our contour
around the circle $|<{\cal A}_3(i)>| = U$. 

Then, the important
parameter for minimizing $V_{eff}$ is $\varphi $, 
and the minimum along the contour occurs at
$\varphi = 0$. The local maximum in $V_{eff}$ lies along the radial direction
$U$ (see figure \ref{effpot2f}),  
which is irrelevent given that  
our variation of the potential is contrained to lie along  
the contour. As $m
\rightarrow 0$, the minimum flattens and the whole contour becomes degenerate. 
By this argument, one will have a dynamically generated non-zero VEV for
${\cal A}_3$, even in the absence of a bare mass term. Although this
could equally be applied to any of the terms in the bi-linear expansion of 
$M_i$, we know 
from the
discussion of Vafa-Witten~\cite{Vafa}, briefly reviewed in section 2,  
that the parity-conserving mass ${\cal A}_3$
is energetically favoured.

\newpage

\section*{Appendix 3: Outline of Strong-Coupling Computational Rules}

We want to calculate the following function on the plaquette

\bea
J_p &=&  \left\{\prod_{i,\mu \in p} \int
dV_{i,\mu} I^{tr}_{0}(2\sqrt{y_{i,\mu}}) trV_{p}trV_{p}^{\dagger}
\right\}
\left\{ \prod_{i,\mu \in p} \int 
dV_{i,\mu} I^{tr}_{0}(2\sqrt{y_{i,\mu}}) \right\}^{-1}, \nn \\
&=& Z_p (Z_p^0)^{-1},
\eea
with
\be
I^{tr}_{0}(2\sqrt{y_{i,\mu}}) = 1 + y_{i,\mu} + \frac{y^2_{i\mu}}{4}
+ \frac{y_{i\mu}^3}{36} + \frac{y_{i\mu}^4}{576}\;\;\;;\;\;\;
y_{i,\mu}=tr(A_{i,\mu}V_{i,\mu})tr(\overline{A}_{i,\mu}V^{\dagger}_{i,\mu})
\;\;.
\ee
Our convention will
be to label the sides of the plaquette 1,2,3,4 and the sites A,B,C,D.
We need to perform the group integrals $\int dV_{i,\mu}$ on each
side of the plaquette. In the case of $Z_p^0$ this will give a product
of polynomials in $O(A\overline{A})$ on each side (\ref{zeroth}).
\bea
Z_0 =& \prod_{i,\mu \in p} \int 
dV_{i,\mu} I^{tr}_{0}(2\sqrt{y_{i,\mu}}) =& (1 + \frac{1}{2} tr(A_1
\overline{A}_1) + \cdots) (1 + \frac{1}{2} tr(A_1
\overline{A}_1)+ \cdots) \times \nn \\
&& \times (1 + \frac{1}{2} tr(A_3
\overline{A}_3) + \cdots) (1 + \frac{1}{2} tr(A_4 \overline{A}_4)+
\cdots) \;\;.
\eea
$Z_p$ will not be
separable into products of simple link-polynomials becasue of the
$trV_ptrV_p^{\dagger}$ which connects the colour indices of the links
around the plaquette in a non-trivial
way. For example there could be a term like $tr(A_1 A_2 A_3 A_4
\overline{A}_4 \overline{A}_3 \overline{A}_2 \overline{A}_1)$ in $Z_p$
which would definitely be absent in $Z_p^0$.

$Z_p$ and $Z_p^0$ are both truncated due to the Grassmann fields
$(\Psi \overline{\Psi})$ contained in the $A,\overline{A}$
functions. So eg. a term like $tr(A_1 \overline{A}_1)^4 tr(A_2
\overline{A}_2)$ will be immediately zero because there are too many
fields at site B between sides 1 and 2. However eg. a term like
$tr(A_1 \overline{A}_1)^4 tr(A_3 
\overline{A}_3)^4$ will survive. 

Alot of the integrals in $Z_p$ will turn out to be equal to
their equivalent values in $Z_p^0$.
In fact the group integrals without at least a term of first order in
$(A\overline{A})$ on {\it every side} will just be equal to their
zeroth order result. This is obvious if we consider eg.

\be
\parbox[c]{3.3in}{$$\int dV_{1}dV_{2}dV_{3}dV_{4}
(tr(AV^\dagger)tr(\overline{A}V))^2 
trV_{p}trV^{\dagger}_{p}$$} 
\parbox[c]{0.3in}{ or }
\parbox[c]{2in}{\psfig{figure=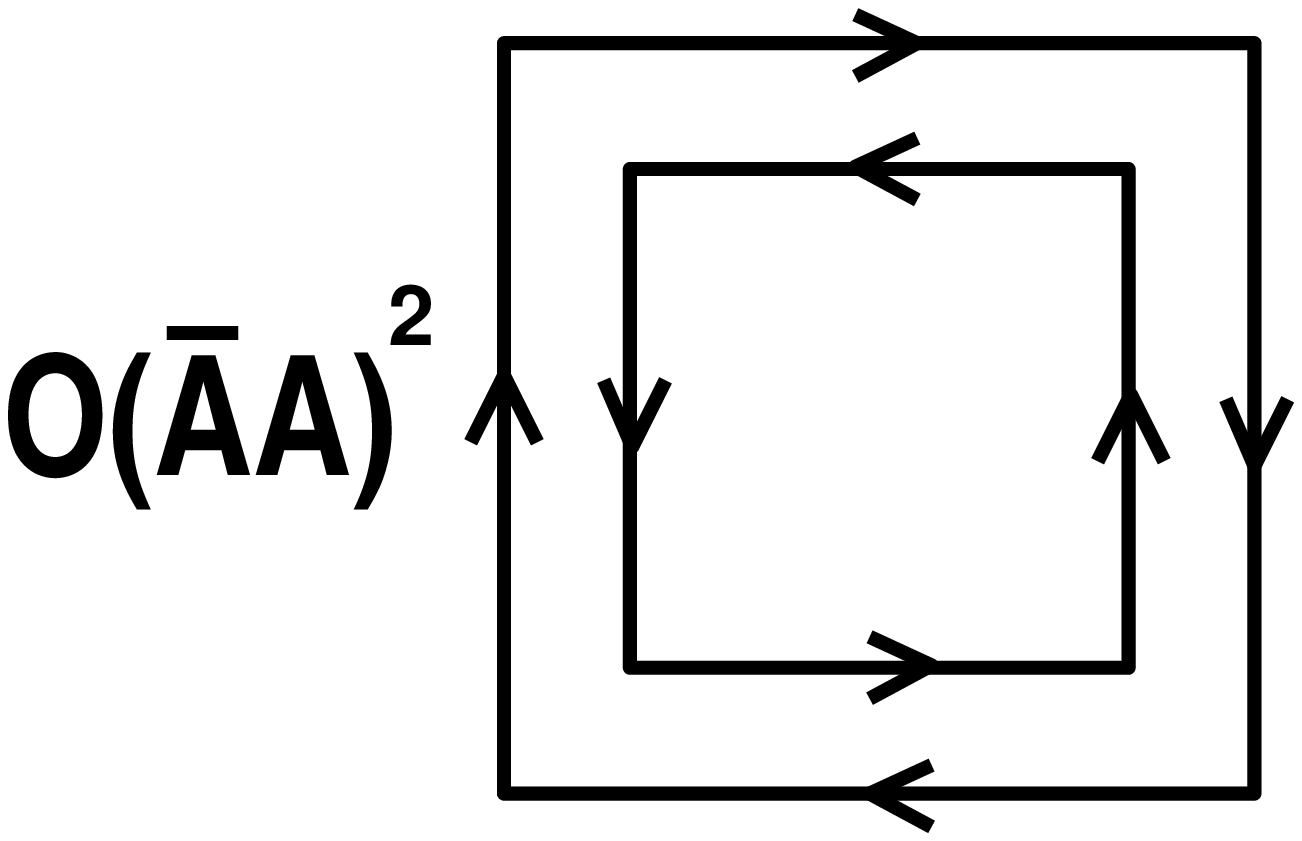,height=1.8cm,width=2.45cm}}\;\;.
\ee
The diagram represents the plaquette with the clockwise flowing arrows
being $trV_p$ and the anti-clockwise $trV_p^{\dagger}$, (the reader
should refer to Creutz \cite{creutz} for a description of diagrammatic
group integration). We do the group integral on one of the blank sides
first. Following 
Creutz \cite{creutz}, this produces delta functions which contract
with the
other group matrices in $trV_ptrV^{\dagger}_p$ all
the way around the plaquette, removing them. This leaves the initial
$(tr(AV^\dagger)tr(\overline{A}V))^2$ without the $trV_ptrV^{\dagger}_p$.

There are only nine terms in $Z_p$ that do not have this
simplification. They are the terms which have at least one power of
$(\overline{A}A)$ on each side, ie.

\bea
&\parbox[c]{1.6in}{\psfig{figure=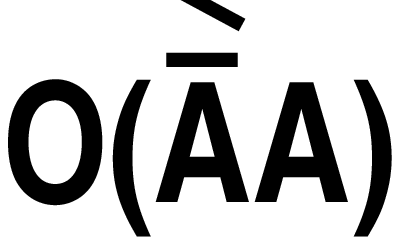,height=1.8cm,width=2.45cm}}
\parbox[c]{0.4in}{$$ + $$}
\parbox[c]{1.6in}{\psfig{figure=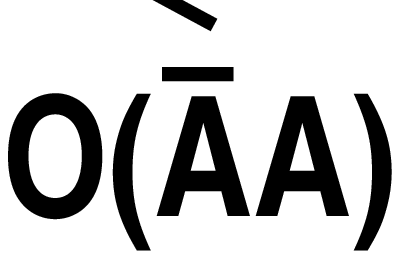,height=1.8cm,width=2.45cm}}
\parbox[c]{0.4in}{$$+$$}
\parbox[c]{1.6in}{\psfig{figure=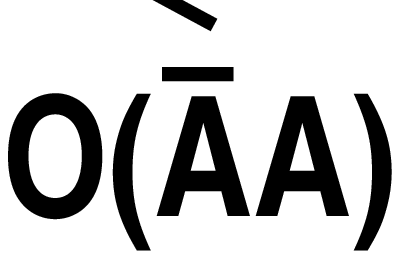,height=1.8cm,width=2.45cm}}
\parbox[c]{0.4in}{$$ +$$}
\parbox[c]{1.6in}{\psfig{figure=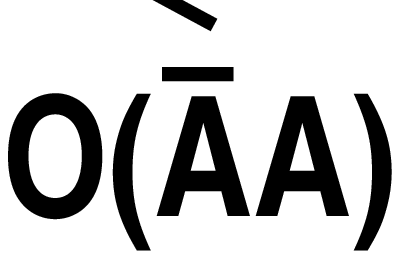,height=1.8cm,width=2.45cm}}
\parbox[c]{0.4in}{$$ +$$}
\parbox[c]{1.6in}{\psfig{figure=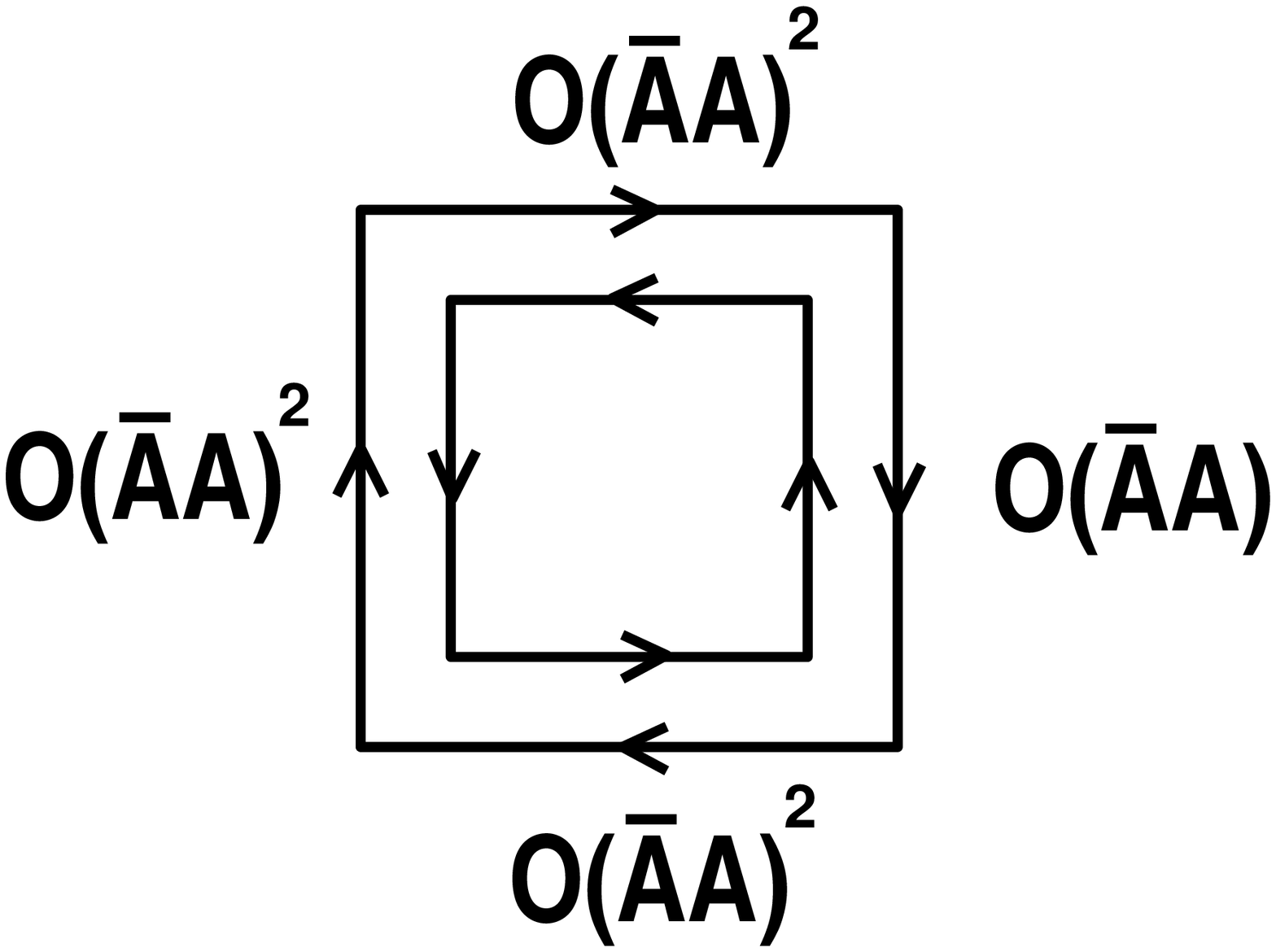,height=1.8cm,width=2.45cm}}&
\nn \\
&\parbox[c]{0.4in}{$$ + $$}
\parbox[c]{1.6in}{\psfig{figure=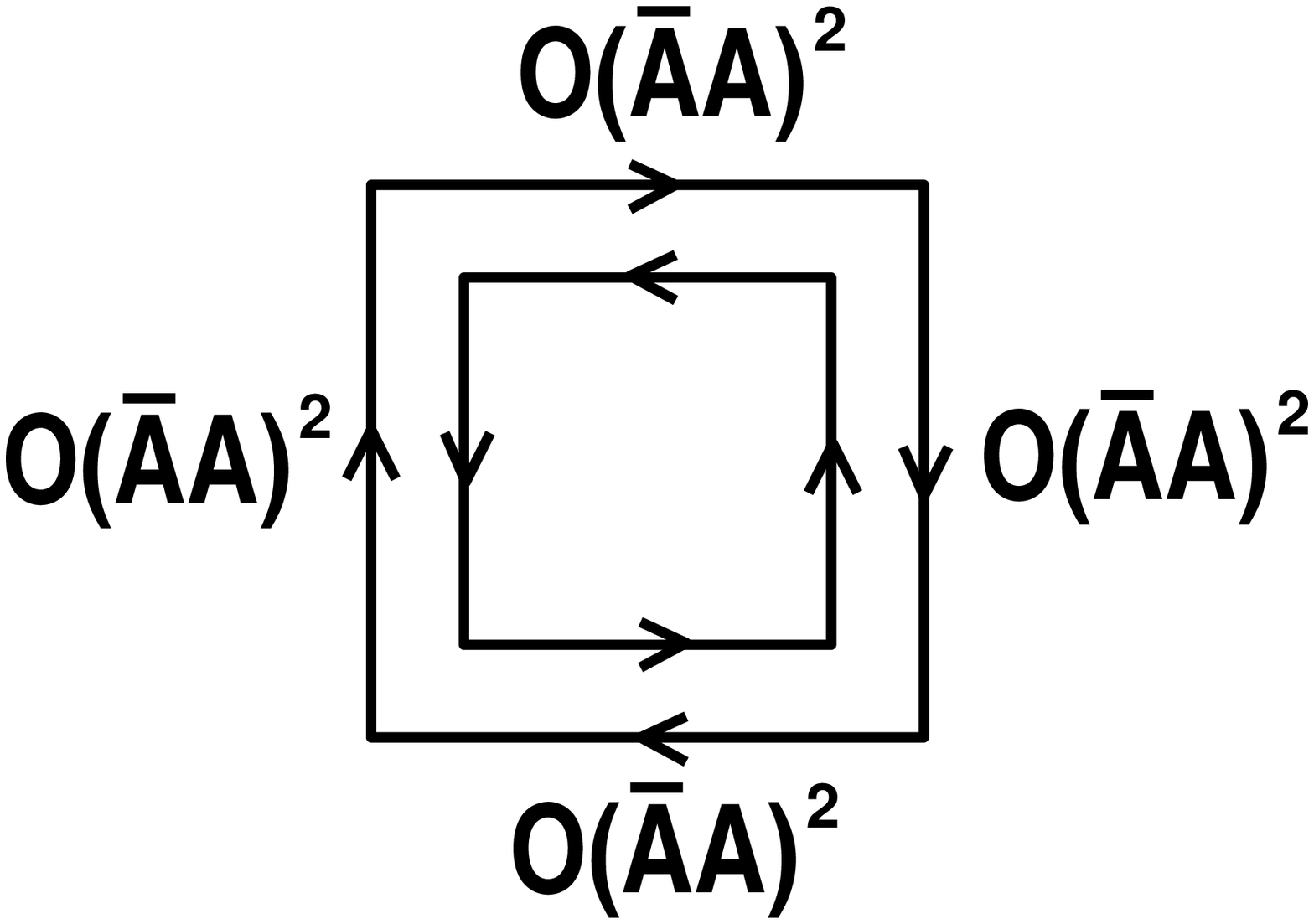,height=1.8cm,width=2.45cm}}
\parbox[c]{0.4in}{$$+$$}
\parbox[c]{1.6in}{\psfig{figure=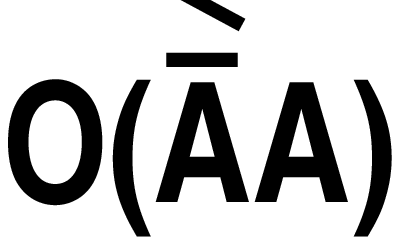,height=1.8cm,width=2.45cm}}
\parbox[c]{0.4in}{$$ +$$}
\parbox[c]{1.6in}{\psfig{figure=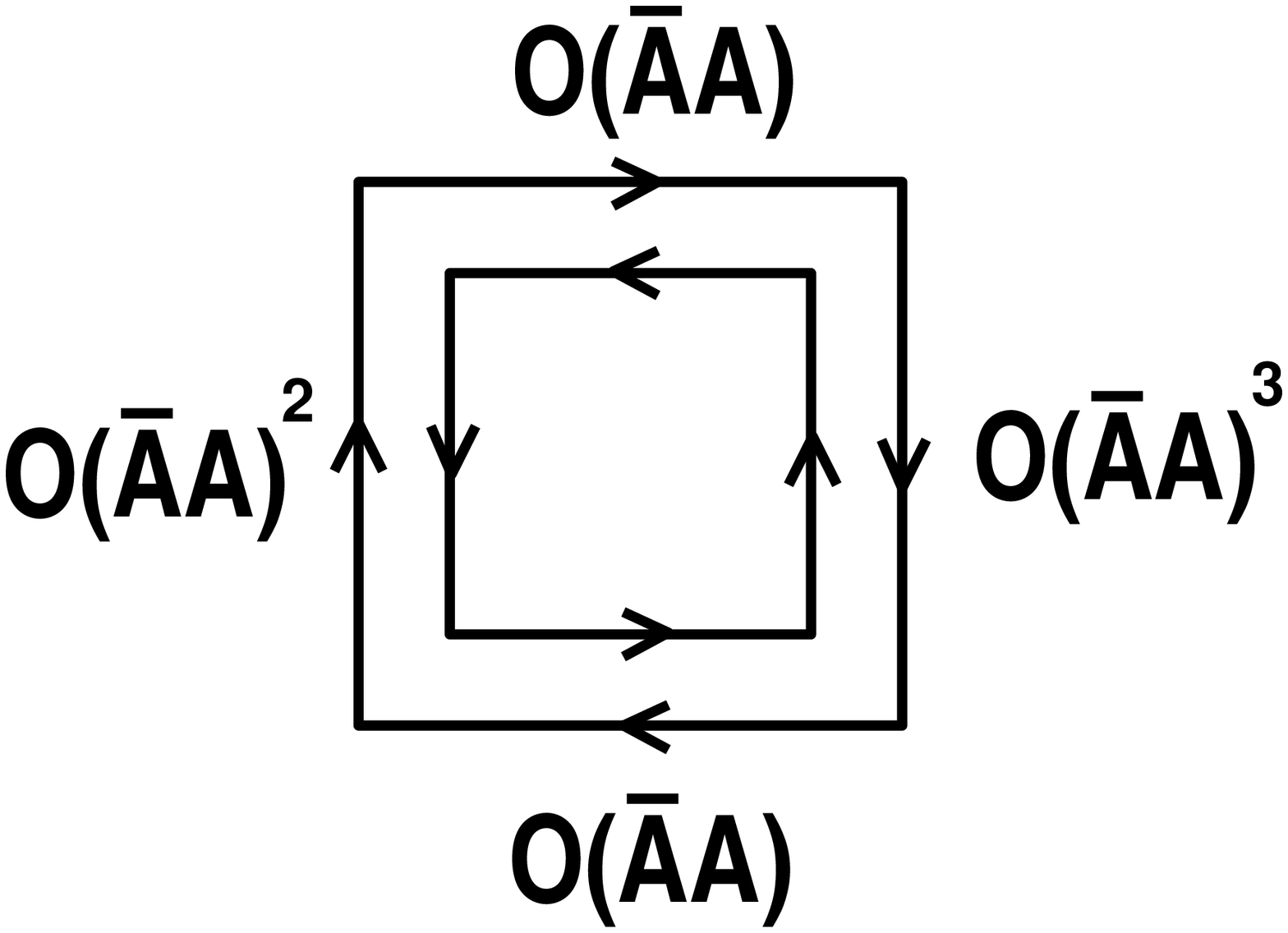,height=1.8cm,width=2.45cm}}
\parbox[c]{0.4in}{$$ +$$}
\parbox[c]{1.6in}{\psfig{figure=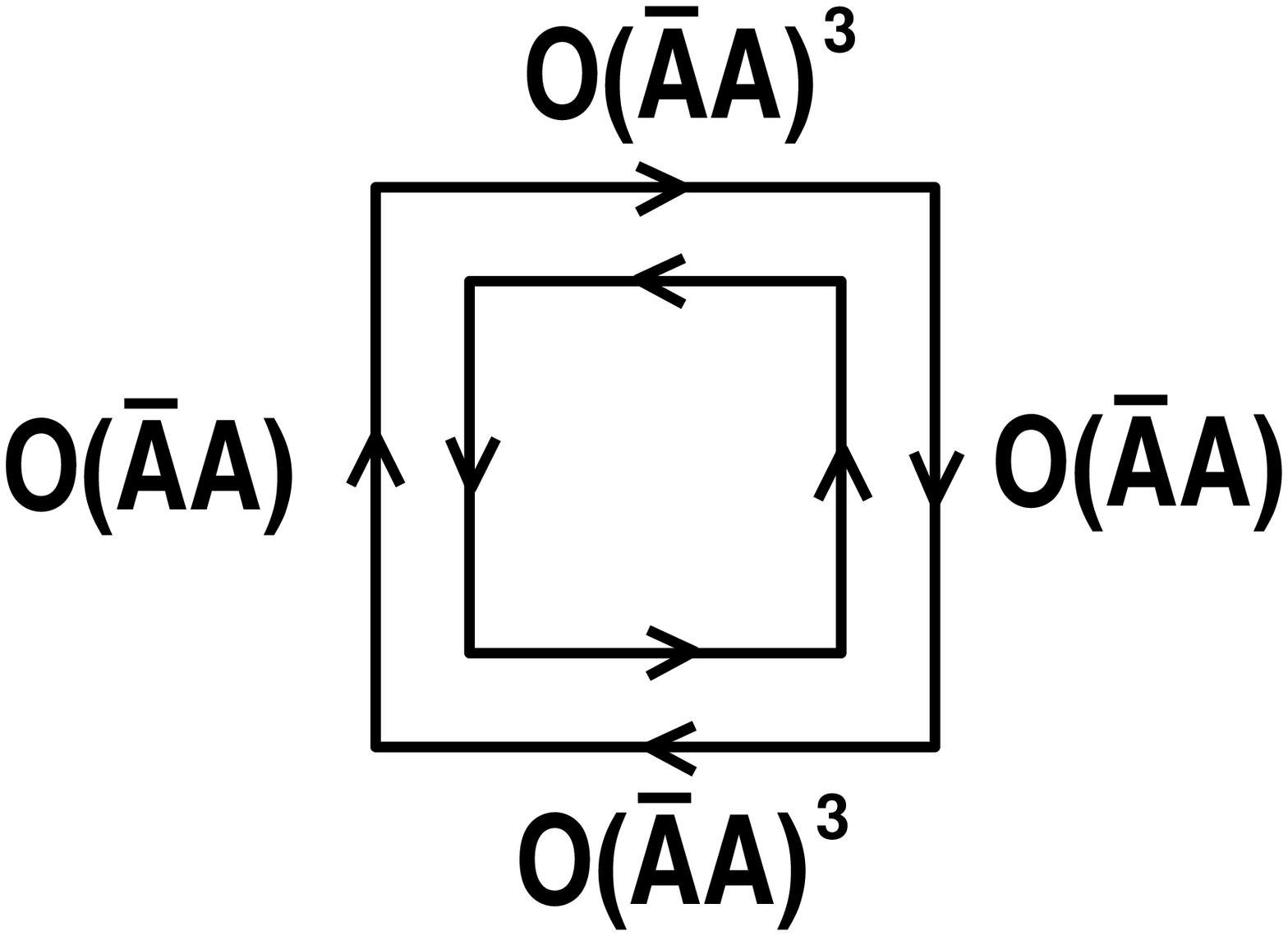,height=1.8cm,width=2.45cm}}\;\;.&
\label{nine}
\eea

The evaluation of these terms starts by spliting up the
$trV_ptrV^{\dagger}_p$ into $V_{ab} \: V_{cd}^{\dagger}$ pairs around the sides
of the plaquette, working out the group 
integrals on each side (following Samuel \cite{sam}) and then
``glueing'' the sides 
together by contracting the spare colour indices (ab,cd). 

As an example we give the explicit form for the first of the integrals in
(\ref{nine})

\bea
&&\int dV_1 dV_2 dV_3 dV_4 \;y_{1}y_{2}y_{3}y_{4} \;trV_{p}trV^{\dagger}_{p}
\nn \\
&=\frac{41}{648}&\;\;tr(\overline{A}_{1}A_{1})tr(\overline{A}_{2}A_{2})tr(\overline{A}_{3}A_{3})tr(\overline{A}_{4}A_{4})
\nn \\
&-\frac{1}{648}&\left\{tr(\overline{A}_{1}A_{1})tr(\overline{A}_{2}A_{2})tr(A_{3}A_{4}\overline{A}_{4}\overline{A}_{3})+tr(\overline{A}_{2}A_{2})tr(\overline{A}_{3}A_{3})tr(A_{4}A_{1}\overline{A}_{1}\overline{A}_{4})\right.
\nn \\
&&\left.+tr(\overline{A}_{3}A_{3})tr(\overline{A}_{4}A_{4})tr(A_{1}A_{2}\overline{A}_{2}\overline{A}_{1})+tr(\overline{A}_{4}A_{4})tr(\overline{A}_{1}A_{1})tr(A_{2}A_{3}\overline{A}_{3}\overline{A}_{2})\right\}
\nn \\
&+\frac{1}{324}&\left\{tr(\overline{A}_{1}A_{1})tr(A_{2}A_{3}A_{4}\overline{A}_{4}\overline{A}_{3}\overline{A}_{2})+tr(\overline{A}_{2}A_{2})tr(A_{3}A_{4}A_{1}\overline{A}_{1}\overline{A}_{4}\overline{A}_{3})\right.
\nn \\
&&\left.+tr(\overline{A}_{3}A_{3})tr(A_{4}A_{1}A_{2}\overline{A}_{2}\overline{A}_{1}\overline{A}_{4})+tr(\overline{A}_{4}A_{4})tr(A_{1}A_{2}A_{3}\overline{A}_{3}\overline{A}_{2}\overline{A}_{1})\right\}
\nn \\
&+\frac{1}{324}&\left\{tr(A_{1}A_{2}\overline{A}_{2}\overline{A}_{1})tr(A_{3}A_{4}\overline{A}_{4}\overline{A}_{3})+tr(A_{4}A_{1}\overline{A}_{1}\overline{A}_{4})tr(A_{2}A_{3}\overline{A}_{3}\overline{A}_{2})\right\}
\nn \\
&-\frac{1}{162}&\left\{tr(A_{1}A_{2}A_{3}A_{4}\overline{A}_{4}\overline{A}_{3}\overline{A}_{2}\overline{A}_{1})+tr(A_{2}A_{3}A_{4}A_{1}\overline{A}_{1}\overline{A}_{4}\overline{A}_{3}\overline{A}_{2})\right.
\nn \\
&&\left.+tr(A_{3}A_{4}A_{1}A_{2}\overline{A}_{2}\overline{A}_{1}\overline{A}_{4}\overline{A}_{3})+tr(A_{4}A_{1}A_{2}A_{3}\overline{A}_{3}\overline{A}_{2}\overline{A}_{1}\overline{A}_{4})\right\}
\nn \\
&+\frac{1}{81}&\;\;tr(A_{1}A_{2}A_{3}A_{4})tr(\overline{A}_{4}\overline{A}_{3}\overline{A}_{2}\overline{A}_{1})\;\;.
\label{longone}
\eea
It it obvious from (\ref{longone}) above that the nine integrals are
not separable into products of link-functions.
We re-write $Z_p = Z_p^0 + Z_p^{'}$ where $Z_p^{'}$ contains the above
nine integrals minus the corresponding integrals in $Z_p^0$ (without
the $trV_ptrV^{\dagger}_p$). So $J_p = 1 +  Z_p^{'} (Z_p^0)^{-1}$.

Expanding $(Z_p^0)^{-1}$ in powers of $O(A\overline{A})$
\bea
(Z_p^0)^{-1} = \prod_{i,\mu \in p}
&\left\{1-\frac{1}{2}tr(\overline{A}_{i,\mu}A_{i,\mu})+
\frac{1}{12}[tr(\overline{A}_{i,\mu}A_{i,\mu})]^2 
+\frac{1}{12}tr[(\overline{A}_{i,\mu}A_{i,\mu})^2]+  \right.\nn \\ 
& \left. - \frac{1}{144}[tr(\overline{A}_{i,\mu}A_{i,\mu})]^3
- \frac{1}{24}tr[(\overline{A}_{i,\mu}A_{i,\mu})^3]
+ \frac{19}{1920} [tr(\overline{A}_{i,\mu}A_{i,\mu})]^4 \right. \nn \\
& \left. - \frac{49}{2880}
[tr(\overline{A}_{i,\mu}A_{i,\mu})]^2tr[(\overline{A}_{i,\mu}A_{i,\mu})^2]
+ \frac{37}{5760} tr[(\overline{A}_{i,\mu}A_{i,\mu})^2]^2 \right\},
\label{invZ}
\eea
where we have truncated up to order $O(A\overline{A})^4$

Looking at example term in $Z_p^{'}$ eg.

\be
\parbox[c]{3.3in}{$$\int dV_{1}dV_{2}dV_{3}dV_{4}
\frac{1}{64} (y_1)^2 (y_2)^2 y_3 (y_4)^2 
trV_{p}trV^{\dagger}_{p}$$} 
\parbox[c]{0.3in}{ or }
\parbox[c]{2in}{\psfig{figure=dots1222.eps,height=1.8cm,width=2.45cm}}\;\;.
\ee
When evaluated this will be an even more complicated function of
$A_1,\overline{A}_1, A_2,\overline{A}_2, A_3,\overline{A}_3,
$ $A_4, \overline{A}_4$ than (\ref{longone}). However although it will
have four pairs of 
$\Psi\overline{\Psi}$ at sites B and C (so two pairs of $MM$) it will
have three pairs of  $\Psi\overline{\Psi}$ at sites A and D which
cannot be arranged solely into $MM$ pairs. In order to give a non-zero
contribution, this diagram will have to be multiplied by the factor
$-\frac{1}{2} tr(A_4\overline{A}_4)$ coming from $(Z_p^0)^{-1}$. This
will then give a diagram 
with $O(A\overline{A})^2$ on each side and all $\Psi\overline{\Psi}$
fields can be re-arranged into $tr(M_iM_i)$ form.
So we must evaluate each of the nine diagrams in (\ref{nine}) and then
multiply each by whatever terms in $(Z^0_p)^{-1}$ (\ref{invZ}) which
will give a contribution containing pairs of $\Psi\overline{\Psi}$ at
each site.

There will be only four ways of arranging the
$O(A\overline{A})^n$ terms on each side of the plaquette to ensure
that on each corner there are either two $\Psi\overline{\Psi}$ pairs
or four $\Psi\overline{\Psi}$ pairs. These are shown below.

\be
\parbox[c]{1.6in}{\psfig{figure=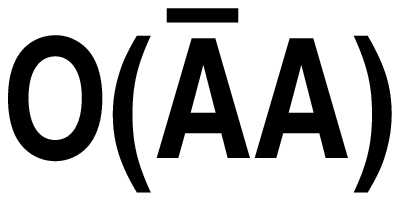,height=1.8cm,width=2.45cm}}
\parbox[c]{0.4in}{$$ or $$}
\parbox[c]{1.6in}{\psfig{figure=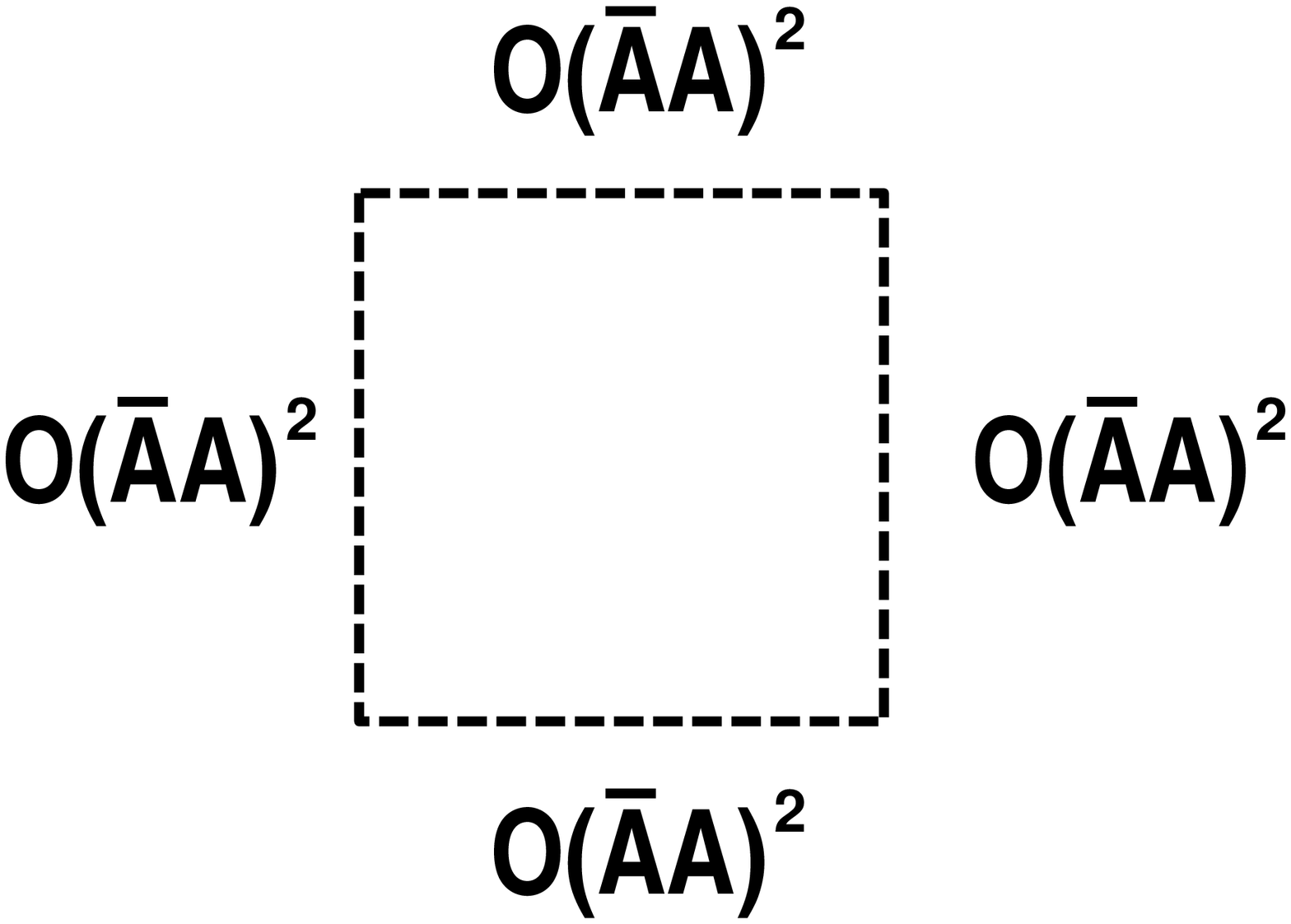,height=1.8cm,width=2.45cm}}
\parbox[c]{0.4in}{$$or$$}
\parbox[c]{1.6in}{\psfig{figure=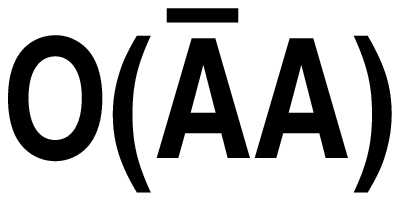,height=1.8cm,width=2.45cm}}
\parbox[c]{0.4in}{$$ or$$}
\parbox[c]{1.6in}{\psfig{figure=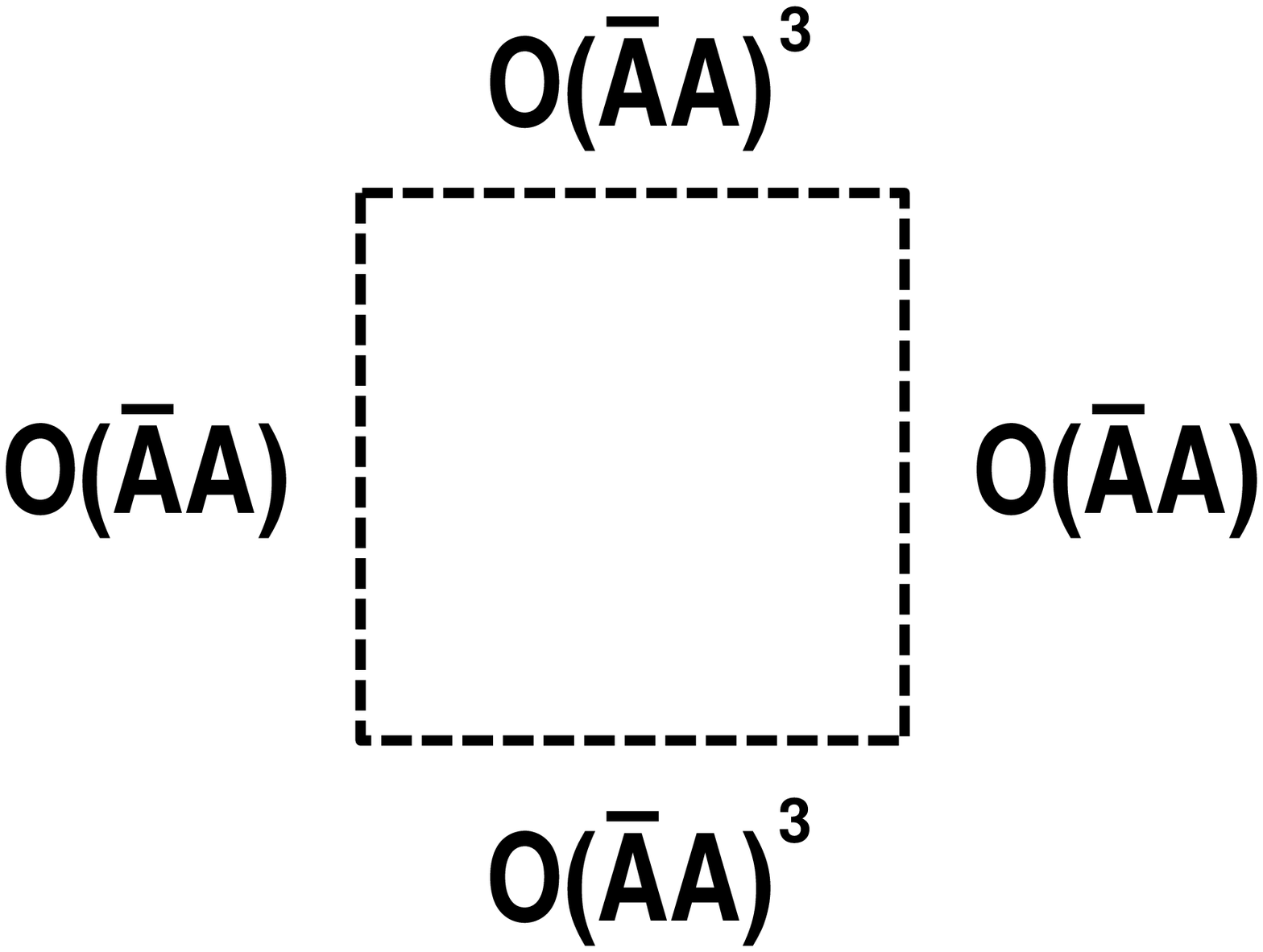,height=1.8cm,width=2.45cm}}\;\;.
\label{four}
\ee
\vspace{1cm}
Finally we describe the procedure for summing over all the delta
functions which arise after substituting $<trM_i^{\alpha\beta}
M_i^{\gamma\delta}> = 2U_i^2 \delta^{\alpha\beta}
\delta^{\gamma\delta}$ at each site.

One term which will occur in $Z_p^{'} (Z_p^0)^{-1}$ is 
\be
\parbox[c]{1.8in}{\psfig{figure=dots5.eps,height=1.8cm,width=2.45cm}},
\label{zero}
\ee
where a simple possibility for $O(A\overline{A})$ is
$tr(A_{i,\mu}\overline{A}_{i,\mu})^2$. Of course there will be much more
complicated terms of $O(A\overline{A})^2$ where the $A,\overline{A}$ are
connected by colour indices to other $A,\overline{A}$ on adjacent
sides, but the procedure will be the same.

\be
[tr(\overline{A}_1A_1)]^2 = 
[\overline{\Psi}_{A}^{b,\alpha}
(-\gamma_\mu)^{\alpha\beta}\Psi_{B}^{a,\beta}
\overline{\Psi}_{B}^{a,\gamma}
(\gamma_\mu)^{\gamma\delta} \Psi_{A}^{b,\delta}]
\times
[\overline{\Psi}_{A}^{d,\epsilon}
(-\gamma_\mu)^{\epsilon\zeta}\Psi_{B}^{c,\zeta}
\overline{\Psi}_{B}^{c,\theta}
(\gamma_\mu)^{\theta\eta} \Psi_{A}^{d,\eta}],
\ee
and this can be rewritten,
\be
\sim (\Psi_{A}^{b,\delta} \overline{\Psi}_{A}^{b,\alpha}
\Psi_{A}^{d,\eta} \overline{\Psi}_{A}^{d,\epsilon}
)(\gamma_\mu^{\alpha\beta} \gamma_\mu^{\gamma\delta}
\gamma_\mu^{\epsilon\zeta} \gamma_\mu^{\theta\eta} 
)(\Psi_{B}^{a,\beta} \overline{\Psi}_{B}^{a,\gamma} \Psi_{B}^{c,\zeta}
\overline{\Psi}_{B}^{c,\theta}),
\ee
or in a more abstract form
\be
a^A.X^{AB}.a^B,
\label{abst}
\ee
where $.$ represents the contraction over the Dirac indices. So all
around the plaquette we have 
$[tr(\overline{A}_1A_1)]^2 [tr(\overline{A}_2A_2)]^2
[tr(\overline{A}_3A_3)]^2 
[tr(\overline{A}_4A_4)]^2$, written in this abstract form as

\be
(a^A.X^{AB}.a^B)(a^B.X^{BC}.a^C)(a^C.X^{CD}.a^D)(a^D.X^{DA}.a^A),
\ee
or
\be
Tr[(a^Aa^A).X^{AB}.(a^Ba^B).X^{BC}.(a^Ca^C).X^{CD}.(a^Da^D).X^{DA}],
\label{one}
\ee
where $Tr$ is, at the moment, just indicating that all Dirac indices
are being summed over. $(a^Aa^A)$ now has 8 Dirac indices, 4 of which
it contracts with $X^{AB}$ and 4 with $X^{DA}$.

The next step, is to re-arrange the 4
$(\Psi_i , \overline{\Psi}_i)$ pairs in $(a^ia^i)$ into 2 meson
pairs. As mentioned in Appendix 1, this can be done in three, equally
likely ways. 
The function at site B, $a^Ba^B$ will be of the form

\be
a^Ba^B \sim 
\Psi_{B}^{a,\beta}
\overline{\Psi}_{B}^{a,\gamma}
\Psi_{B}^{c,\zeta}
\overline{\Psi}_{B}^{c,\theta}
\Psi_{B}^{f,\rho}
\overline{\Psi}_{B}^{f,\kappa}
\Psi_{B}^{h,\chi}
\overline{\Psi}_{B}^{h,\sigma},
\label{abab}
\ee
where $\Psi_{B}^{a,\beta} \overline{\Psi}_{B}^{a,\gamma}
\Psi_{B}^{c,\zeta} \overline{\Psi}_{B}^{c,\theta}$ comes from side AB
and $\Psi_{B}^{f,\rho} \overline{\Psi}_{B}^{f,\kappa}
\Psi_{B}^{h,\chi} \overline{\Psi}_{B}^{h,\sigma}$ from side BC. We
rewrite this as

\bea
&=\frac{1}{3} \Psi_{B}^{a,\beta}
\overline{\Psi}_{B}^{c,\theta}
\Psi_{B}^{c,\zeta}
\overline{\Psi}_{B}^{a,\gamma} 
\Psi_{B}^{f,\rho}
\overline{\Psi}_{B}^{h,\sigma}
\Psi_{B}^{h,\chi}
\overline{\Psi}_{B}^{f,\kappa} +
\frac{1}{3} \Psi_{B}^{a,\beta}
\overline{\Psi}_{B}^{h,\sigma}
\Psi_{B}^{h,\chi}
\overline{\Psi}_{B}^{a,\gamma}
\Psi_{B}^{f,\rho}
\overline{\Psi}_{B}^{c,\theta}
\Psi_{B}^{c,\zeta}
\overline{\Psi}_{B}^{f,\kappa}& \nn \\
&+ \frac{1}{3} \Psi_{B}^{a,\beta}
\overline{\Psi}_{B}^{f,\kappa}
\Psi_{B}^{f,\rho}
\overline{\Psi}_{B}^{a,\gamma}
\Psi_{B}^{c,\zeta}
\overline{\Psi}_{B}^{h,\sigma}
\Psi_{B}^{h,\chi}
\overline{\Psi}_{B}^{c,\theta} \;\;.&
\eea
Now we substitute in $<M_i^{ab,\alpha\beta}> = U_i \sigma_3^{ab}
\delta^{\alpha\beta}$ to get

\be
\frac{4}{3}U_B^4\left\{\delta^{\beta\theta}\delta^{\zeta\gamma}
\delta^{\rho\sigma} \delta^{\chi\kappa} + \delta^{\beta\sigma} 
\delta^{\chi\gamma} \delta^{\rho\theta} \delta^{\zeta\kappa} +
\delta^{\beta\kappa} \delta^{\rho\gamma} \delta^{\zeta\sigma}
\delta^{\chi\theta} \right\}\;\;.
\label{delta}
\ee

So we have effectively replaced $(a^ia^i)$ with $\pm\frac{4}{3} U_i^4
\Delta^{\beta\theta\zeta\gamma\rho\sigma\chi\kappa}$, with $\Delta$
representing the sum of products of delta fns (the bit in \{\}) ,
which has 8 Dirac
indices. The $\pm$ above appears because when we rearrange the
$(\Psi_i , \overline{\Psi}_i)$  we need to be careful about the
anti-commuting Grassmann numbers.

Since each Dirac index has 2 possible values (it's a 2x2
representation of the Dirac algebra) and
half the indices contract to the right and half to the left, we can
re-represent the function $\Delta$ (a $2^8$ tensor) with a 16x16
matrix. In (\ref{delta}) the four indices $\beta \theta \zeta \gamma$
connect via four $\gamma$ matrices to site A, and $\rho \sigma \chi
\kappa$ connect via four $\gamma$ matrices to site C. We can therefore
replace each set of four indices (four pairs = sixteen combinations)
with a single index taking 16 values. So
$\Delta^{(\beta\theta\zeta\gamma)(\rho\sigma\chi\kappa)} =
\Delta^{\Theta\Omega}$.
The same goes for the functions $X$ representing the four gamma
matrices in (\ref{one}) and we
merely need to calculate products of 16x16 matrices around the
plaquette. The $Tr$ in 
(\ref{one}) then becomes a matrix trace. 
Actually working out the matrix form for each product of delta
functions then becomes a matter of book-keeping.

A helpful result is that, because of the symmetry of the problem and the
even number of gamma matrices, we can actually replace the $X$s with
identity matrices.

We have outlined the procedure for working out one term, where all the
corner functions were of the form $(a^ia^i)$ but the same
goes for each term in $Z_p(Z_p^0)^{-1}$. In the case where there are spare
colour indices to contract at the corners we must do this first and
then re-arrange the $\Psi\overline{\Psi}$, but each corner will
eventually be written like (\ref{abab}).

For the other diagrams in (\ref{four}) there will not be four gamma
matrices on the sides but rather two (for $O(A\overline{A})$) or six
(for $O(A\overline{A})^3$) and these can be rewritten as $4 \times 4$
or $64 \times 64$ matrices respectively. So the corner function will
be a $4 \times 4$ matrix for a corner connecting
two $O(A\overline{A})$ sides, and for a corner
connecting a $O(A\overline{A})$ side and a $O(A\overline{A})^3$ side a
$4 \times 64$ matrix (with the transpose $64 \times 4$ for an adjacent
corner). The general procedure however, is identical in each case.
Our computation is given in more detail in \cite{danthes} and is
summarized below.

\noindent
1. Work out the group integrals on each side, the result will be a sum
of terms like (\ref{abst}) \\
2. Write the integral around the whole plaquette as a sum of terms
each of the form (\ref{one}) contracting all spare colour indices. \\
3. Replace each corner functions with the sum of delta functions, being
careful to keep track of $\pm$ due to the anti-commuting nature of $(\Psi_i , \overline{\Psi}_i)$ \\
4. Rewrite the $\Delta$ functions as $16 \times 16$, $4 \times 4$, or $4
\times 64$ matrices \\
5. Trace over the matrices \\

\end{document}